\documentclass[10pt,twocolumn,letterpaper]{article}
\pdfoutput=1

\usepackage{usenix,endnotes}
\usepackage{times}
\usepackage{xcolor}
\usepackage{graphicx,epsfig}
\usepackage[caption=false]{subfig}
\usepackage{titlesec}
\usepackage{float}
\usepackage{varwidth}
\usepackage{multirow}
\usepackage[super]{nth}
\usepackage{algorithm}
\usepackage[noend]{algpseudocode}
\usepackage{array}
\usepackage{placeins}
\usepackage{authblk}
\usepackage{hyperref} 
\usepackage[capitalize]{cleveref}

\makeatletter
\algrenewcommand\ALG@beginalgorithmic{\small}
\makeatother
\definecolor{light-gray}{gray}{0.90}

\titleformat*{\section}{\large\bfseries}
\titleformat*{\subsection}{\normalsize\bfseries}
\titleformat*{\subsubsection}{\normalsize\bfseries}
\titleformat*{\paragraph}{\normalsize\bfseries}
\titleformat*{\subparagraph}{\normalsize\bfseries}

\titlespacing*{\section}{0pt}{12pt plus 4pt minus 2pt}{6pt plus 2pt minus 2pt}
\titlespacing*{\subsection}{0pt}{12pt plus 4pt minus 2pt}{4pt plus 2pt minus 2pt}

\begin{document}

\title{\Large \bf Don't cry over spilled records: Memory elasticity of data-parallel applications and its application to cluster scheduling}
\author[*]{Calin Iorgulescu}
\author[*]{Florin Dinu}
\author[$\ddagger$]{Aunn Raza}
\author[$\dagger$]{Wajih Ul Hassan}
\author[*]{Willy Zwaenepoel}

\makeatletter
\renewcommand\AB@affilsepx{\hspace*{1cm}\protect\Affilfont}
\affil[*]{EPFL}
\affil[$\dagger$]{UIUC}
\affil[$\ddagger$]{NUST Pakistan}

\date{}
\maketitle

\begin{abstract}
Understanding the performance of data-parallel workloads when
resource-constrained has significant practical importance but unfortunately has
received only limited attention. This paper identifies, quantifies and
demonstrates memory elasticity, an intrinsic property of data-parallel tasks.
Memory elasticity allows tasks to run with significantly less memory that they
would ideally want while only paying a moderate performance penalty.  For
example, we find that given as little as 10\% of ideal memory,
PageRank and NutchIndexing Hadoop reducers become only 1.2x/1.75x and
1.08x slower. We show that memory elasticity is prevalent in the Hadoop, Spark,
Tez and Flink frameworks.  We also show that memory elasticity is predictable
in nature by building simple models for Hadoop and extending them to Tez and
Spark.

To demonstrate the potential benefits of leveraging memory elasticity, this
paper further explores its application to cluster scheduling. In this setting,
we observe that the resource vs. time trade-off enabled by memory elasticity
becomes a task queuing time vs task runtime trade-off. Tasks may complete
faster when scheduled with less memory because their waiting time is reduced.
We show that a scheduler can turn this task-level trade-off into improved job
completion time and cluster-wide memory utilization. We have integrated memory
elasticity into Apache YARN. We show gains of up to 60\% in average job
completion time on a 50-node Hadoop cluster. Extensive simulations show similar
improvements over a large number of scenarios.

\end{abstract}

\section{Introduction}

The recent proliferation of data-parallel
workloads~\cite{spark-nsdi12,vldb2012-workloads,googletrace-socc} has made
efficient resource management~\cite{heracles, borg, apollo} a top priority in today's computing clusters. A
popular approach is to better estimate workload resource needs to avoid
resource wastage due to user-driven over-estimations~\cite{borg, quasar, morpheus}.
Another is to over-commit server resources to cope with the variability of
workload resource usage~\cite{borg, apollo, rayon}. Unfortunately, only a few
efforts~\cite{quasar} have touched on the malleability of data-parallel
workloads when resource-constrained. The study of malleability is complementary
to solutions for over-estimations and variability. While the latter two attempt
to accurately track the actual workload resource usage, the former is about
allocating to applications fewer server resources than they would ideally need.
A thorough understanding of the trade-offs involved in resource malleability is
useful in many contexts ranging from improving cluster-wide resource efficiency
and cluster provisioning to reservation sizing in public clouds, disaster
recovery, failure recovery and cluster scheduling.

The main contribution of this paper is identifying, quantifying and
demonstrating memory elasticity, an intrinsic property of data-parallel
workloads. We define memory elasticity as the property of a data-parallel task
to execute with only a moderate performance penalty when memory-constrained.
Memory elasticity pertains to tasks involved in data shuffling operations.
Data shuffling is
ubiquitous~\cite{shufflewatcher,netw-aware-sched-sig15,shuffling-nsdi12}. It is
required, in all data-parallel frameworks, to implement even the simplest
data-parallel applications. Thus, most tasks are involved in shuffling and show
memory elasticity: mappers and reducers in MapReduce, joins and by-key
transformations (reduce, sort, group) in Spark, and mappers, intermediate and
final reducers in Tez. 

Despite significant differences in the designs of popular data-parallel
frameworks, shuffling operations share across these frameworks a common,
tried-and-tested foundation in the use of merge-sort algorithms that may also
use secondary storage~\cite{cloudera-shuffle}. The memory allocated to a task
involved in shuffling has a part for shuffling and a part for execution. The
best task runtime is obtained when the shuffle memory is sized such that all
shuffle data fits in it. This allows the shuffle to perform an efficient
in-memory-only merge-sort. If the shuffle memory is insufficient, an external
merge-sort algorithm is used. 

The key insight behind memory elasticity is that under-sizing shuffle memory
can lead to considerable reductions in task memory allocations at the expense
of only moderate increases in task runtime. Two factors contribute to the
sizeable memory reductions. First, shuffle memory is usually a very large
portion of the task memory allocation (70\% by default in Hadoop). Second,
external merge-sort algorithms can run with very little memory because they can
compensate by using secondary storage. A couple of factors also explain why the
task runtime increases only moderately when shuffle memory is under-sized.
First, a data-parallel task couples shuffling with CPU-intensive processing
thus making far less relevant the performance gap between external and
in-memory merge-sort. Second, disk accesses are efficient as the disk is
accessed sequentially.  Third, the performance of external merge-sort
algorithms remains stable despite significant reductions in shuffle memory (a
k-way merge is logarithmic in k). 

Thus, memory elasticity presents an interesting resource vs. time trade-off.
This paper quantifies this trade-off and its implications using extensive
experimental studies. We find that memory elasticity is prevalent across the
Hadoop, Spark, Tez and Flink frameworks and across several popular workloads.
In all cases, the performance penalty of memory elasticity was moderate despite
sizeable reductions in task memory allocations.  Let M be the task memory
allocation that minimizes task runtime by ensuring that all shuffle data fits
in shuffle memory. Given as little as 10\% of M, PageRank and
NutchIndexing Hadoop reducers become only 1.22x/1.75x and 1.08x slower.
For Hadoop mappers the largest encountered penalty is only 1.5x. For Spark, Tez
and Flink the penalties were similar to Hadoop. Furthermore, we show the predictable
nature of memory elasticity which is key to leveraging it in practice. We build
simple models for Hadoop that can accurately describe the resource vs. time trade-off.  
With only small changes the same models apply to Spark and Tez.

To demonstrate the potential benefits of leveraging memory elasticity, this
paper further explores its application to cluster scheduling. Current clusters
host concurrently a multitude of jobs each running a multitude of tasks. In
this setting, we observe that the resource vs. time trade-off of memory
elasticity becomes a task queueing time vs task runtime trade-off. A task
normally has to wait until enough memory becomes available for it but if it is
willing to execute using less memory it might have to wait much less or not at
all. Since the completion time of a task is the sum of waiting time plus
runtime, a significant decrease in waiting time may outweigh an increase in
runtime due to elasticity and overall lead to faster task completion times.  We
show that a scheduler can turn this task-level trade-off into improved job
completion time and improved cluster-wide memory utilization by better packing
tasks on nodes with respect to memory. Scheduling using memory elasticity is an
NP-hard problem because it contains as a special case NP-hard variants of the
RCPSP problem~\cite{rcpsp}, a well-known problem in operations research. We
propose a simple heuristic and show it can yield important benefits: the tasks
in a job can leverage memory elasticity only if that does not lead to a
degradation in job completion time.

We have integrated the concepts of memory elasticity into Apache YARN. On a
50-node Hadoop cluster, leveraging memory elasticity results in up to 60\%
improvement in average job completion time compared to stock YARN. Extensive
simulations show similar improvements over a large number of scenarios.

\section{Memory elasticity in real workloads}
\label{sec:study}

This section presents an extensive study of memory elasticity. We make a few
key points.  First, memory elasticity is generally applicable to several
frameworks and workloads. Our measurements have an emphasis on Hadoop but also
show that elasticity applies to Apache Spark, Tez and Flink. Second, memory
elasticity costs little. The performance degradation due to using elasticity
was moderate in all experiments. Third, elasticity has a predictable nature and
thus can be readily modeled. We provide a model for Hadoop and with only simple
changes apply it to Tez and a Spark Terasort job. We also detail the causes and
implications of memory elasticity. 

We use the term spilling to disk to refer to the usage of secondary storage by
the external merge-sort algorithms. We call a task under-sized if its memory
allocation is insufficient to avoid spilling to disk during shuffling. We call
a task well-sized otherwise. We call ideal memory the minimum memory allocation
that makes a task well-sized and ideal runtime the task runtime when allocated
ideal memory. We use the term penalty to refer to the performance penalty due
to memory elasticity for under-sized tasks.

\begin{figure*}[t] 
\subfloat[Elasticity for Hadoop mappers\label{fig:hadoop-mapper-elasticity}]
{\includegraphics[width=.23\textwidth,angle=270]{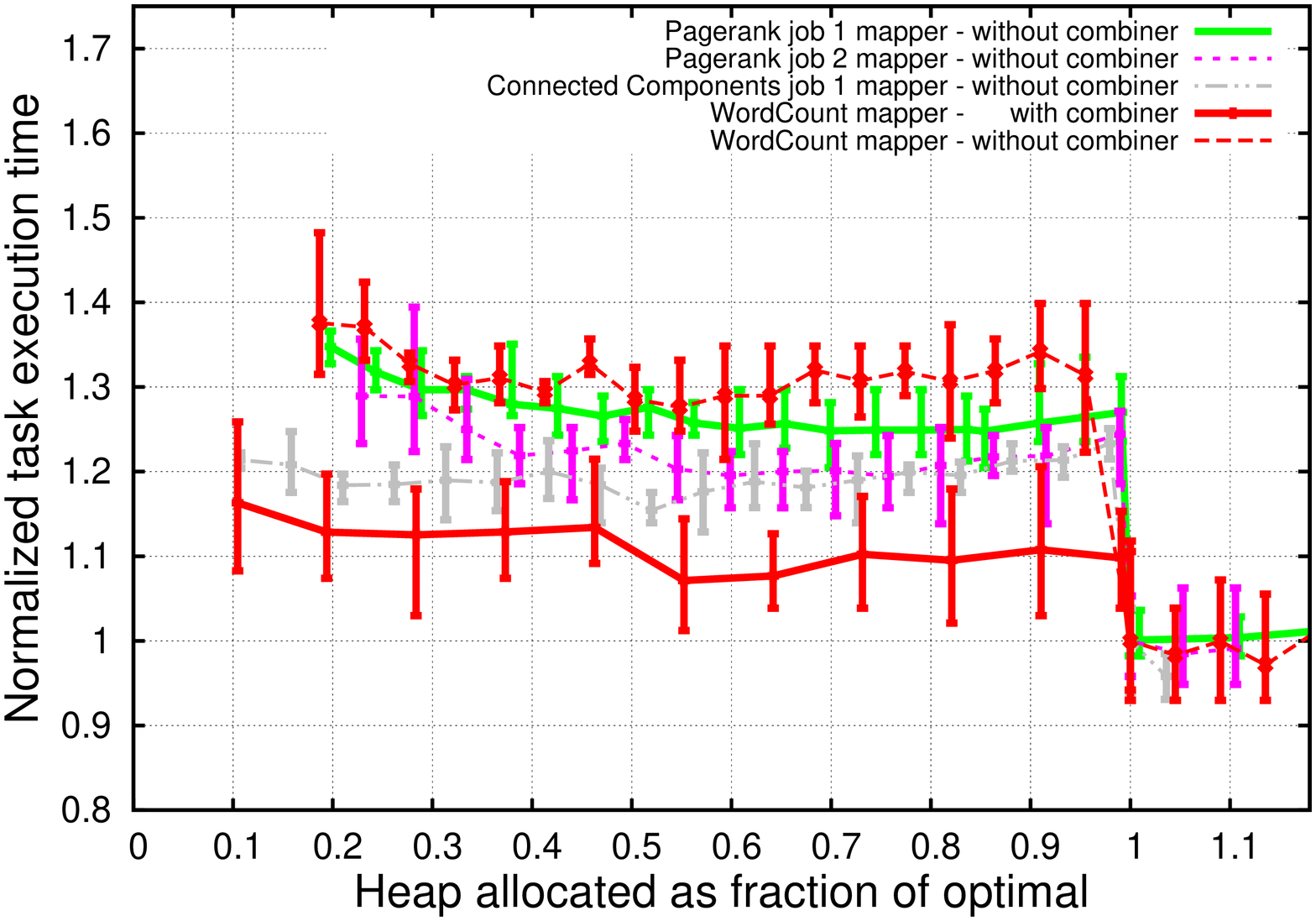}}
\hfill
\subfloat[Elasticity for Hadoop reducers\label{fig:hadoop-elasticity}]
{\includegraphics[width=.23\textwidth,angle=270]{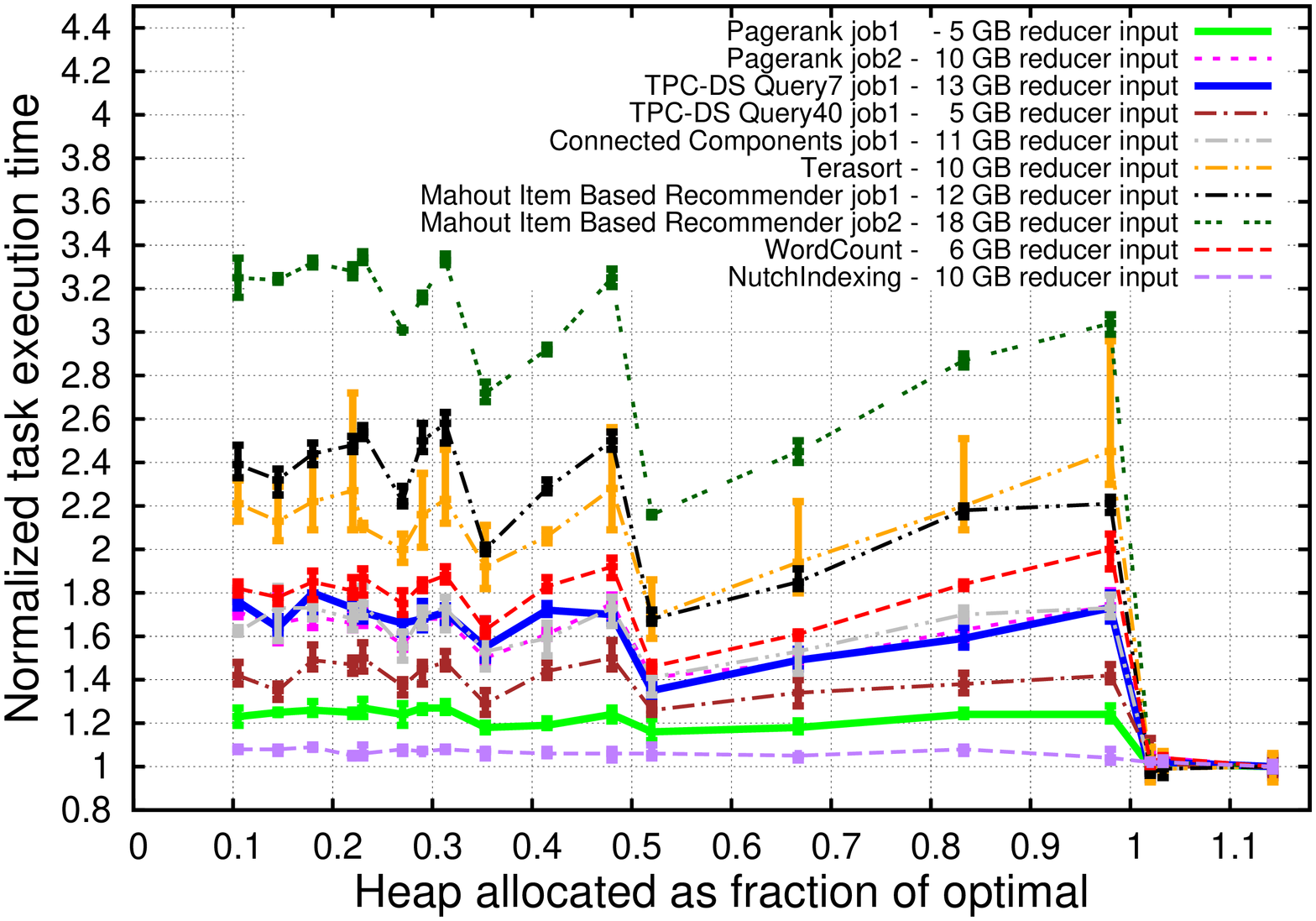}}
\hfill
\subfloat[Accuracy of modeling reducer behavior\label{fig:hadoop-reducer-model}]
{\includegraphics[width=.23\textwidth,angle=270]{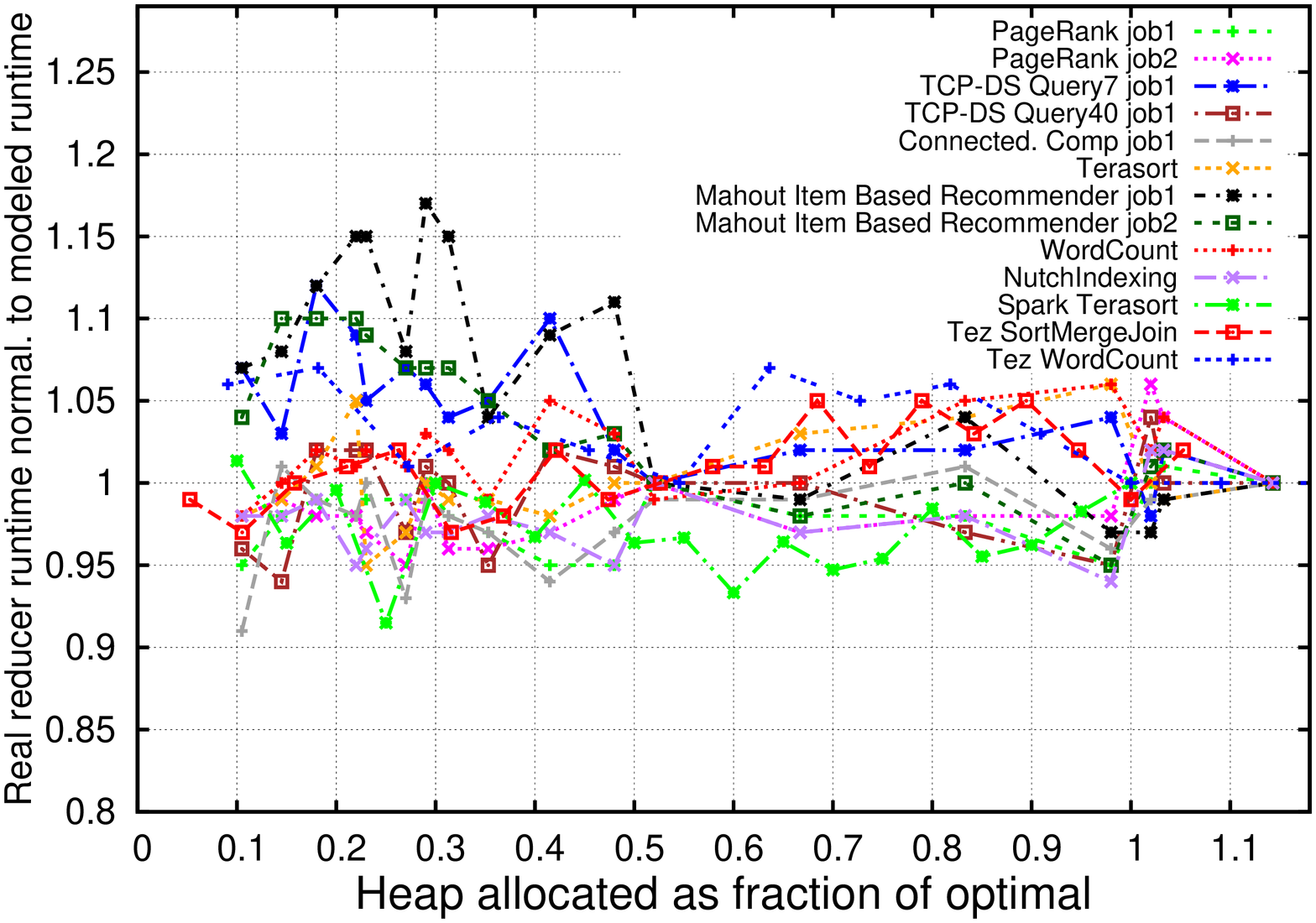}}
\caption{Memory elasticity profiles for Hadoop mappers (a) and reducers (b). Accuracy of our model (c).} 
\end{figure*}

\subsection{Measurement methodology}
For Hadoop we profiled 18 jobs across 10 different applications, most belonging
to the popular HiBench big-data benchmarking suite~\cite{hibench}. The jobs
range from graph processing (Connected Components, PageRank) to web-indexing
(Nutch), machine learning (Bayesian Classification, Item-Based Recommender),
database queries (TPC-DS) and simple jobs (WordCount, TeraSort). For Spark we
profiled TeraSort and WordCount, for Tez we profiled WordCount and SortMerge
Join and for Flink we profiled WordCount.  We used Hadoop 2.6.3, Spark 2.0.0,
Tez 0.7.0 and Flink 1.0.2. However, the same behavior appears in Spark versions
prior to 2.0.0 and Hadoop versions at least as old as 2.4.1 (June 2014).

For accurate profiling we made sure that the profiled task is not collocated
with any other task. To measure the worst case penalties for under-sized tasks
we ensure that disk I/O operations for spills actually go to the drive and not
to the OS buffer cache. For this, we ran each task in a separate Linux cgroups
container.  We minimize the amount of buffer cache available to a task by
setting the cgroups limits as close to the JVM heap size as possible. As an
alternative we also modified Hadoop to perform disk spills using direct I/O
thus bypassing completely the OS buffer cache. The two solutions gave
consistently similar results.

\subsection{Memory elasticity for Hadoop mappers}

Elasticity for mappers occur on their output side.  The key-value pairs output
by map function calls are written to an in-memory buffer. If the mapper is
well-sized then the buffer never fills up. In this case, when the mapper
finishes processing its input, the buffer contents are written to disk into one
sorted and partitioned file (one partition per reducer).  If the mapper is
under-sized, the buffer fills up while the mapper is still executing map
function calls. The buffer contents are spilled to disk and the buffer is
reused. For under-sized mappers, at the end there is an extra merge phase that
merges together all existing spills. If combiners are defined then they are
applied before spills.

\vspace{0.01in} \noindent{\bf The impact of elasticity on mapper
runtime\hspace{0.2in}} \cref{fig:hadoop-mapper-elasticity} shows the
mapping between normalized mapper runtime (y-axis) and allocated heap size
(x-axis) for several Hadoop mappers. We call this mapping the memory elasticity
profile. The penalties are moderate. For example, an under-sized WordCount
mapper is about 1.35x slower than when well-sized. If the same mapper uses a
combiner, then the penalty is further reduced (1.15x) because less data is
written to disk. The maximum encountered penalty across all mappers is 1.5x.

\vspace{0.01in} \noindent{\bf Why penalties are not larger\hspace{0.2in}} 
As explained in the introduction, three factors limit the penalties.  First,
the mapper couples shuffling with CPU-intensive work done by map
function calls. Second, disk accesses are efficient as the disk is accessed
sequentially. Third, the performance of external merge-sort algorithms remains
stable despite significant reductions in shuffle memory.

\vspace{0.01in} \noindent{\bf The shape of the memory elasticity
profile\hspace{0.2in}} The elasticity profile of a mapper resembles a step
function. The reason is that under-sized mappers perform an extra merge phase
which takes a similar amount of time for many different under-sized
allocations.

\vspace{0.01in} \noindent{\bf Modeling memory elasticity for mappers\hspace{0.2in}}
A step function is thus a simple and good approximation for modeling memory
elasticity for Hadoop mappers. To build this model two training runs are
needed, one with an under-sized mapper and one with a well-sized mapper.  The
runtime of the under-sized mapper can then be used to approximate the mapper
runtime for any other under-sized memory allocations.

\begin{figure*}[t] 
\subfloat[Elasticity for Spark, Tez and Flink reducers\label{fig:sparktez-elasticity}]
{\includegraphics[width=.23\textwidth,angle=270]{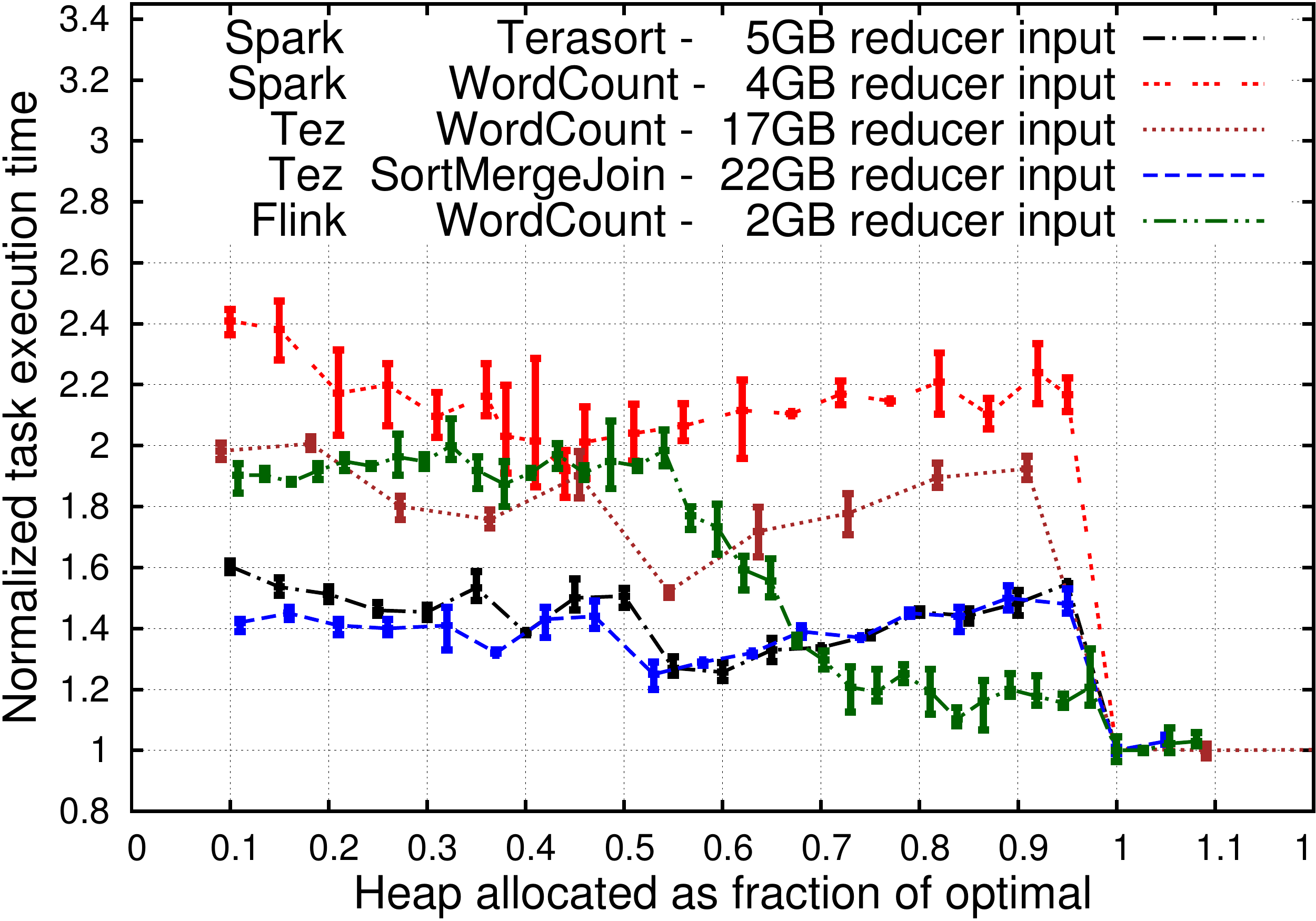}}
\hfill
\subfloat[Spilling vs paging\label{fig:paging_vs_spilling}]
{\includegraphics[width=.23\textwidth, angle=270]{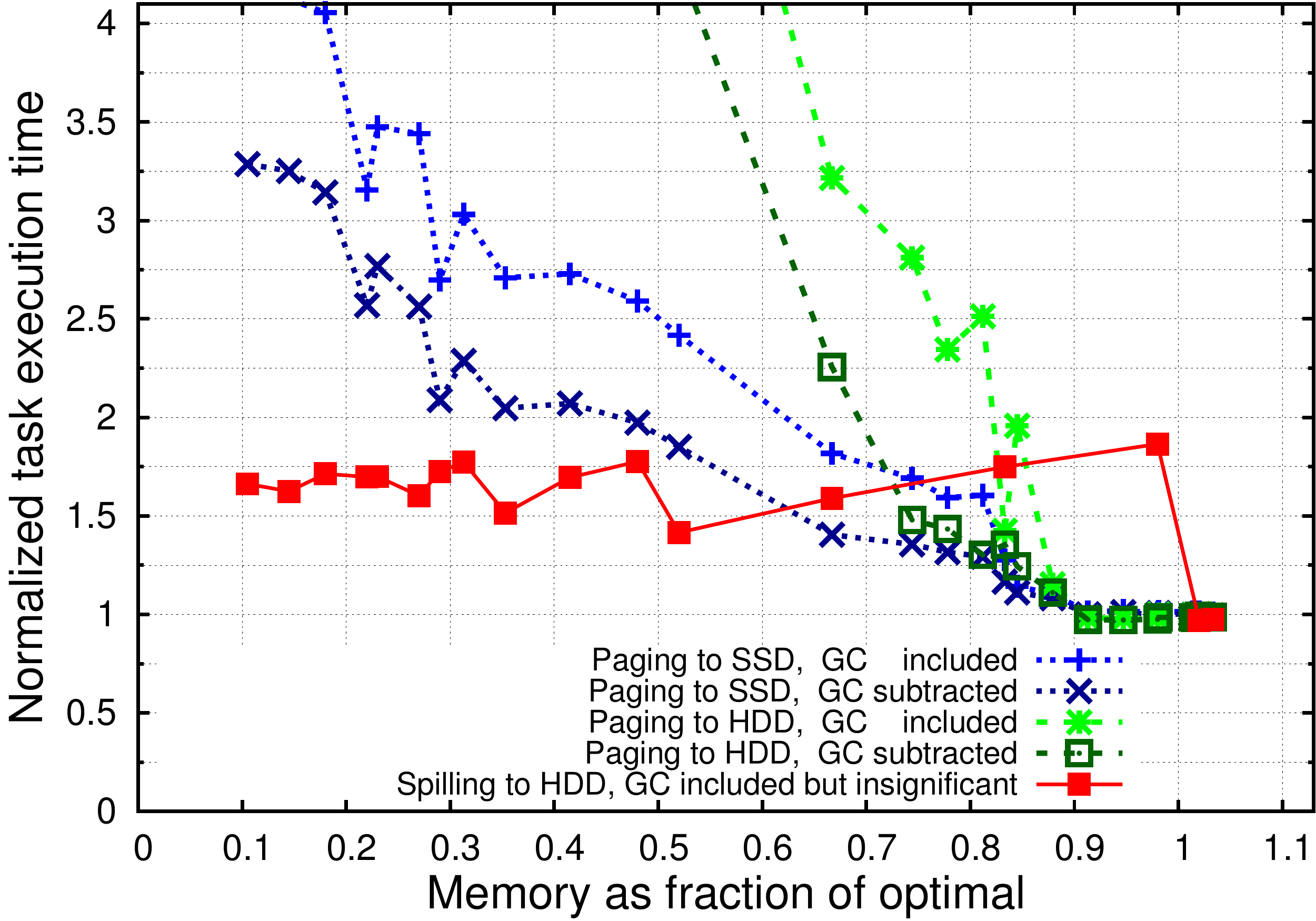}}
\hfill
\subfloat[Impact of contention on Hadoop reducers\label{fig:concurrency}]
{\includegraphics[width=.23\textwidth, angle=270]{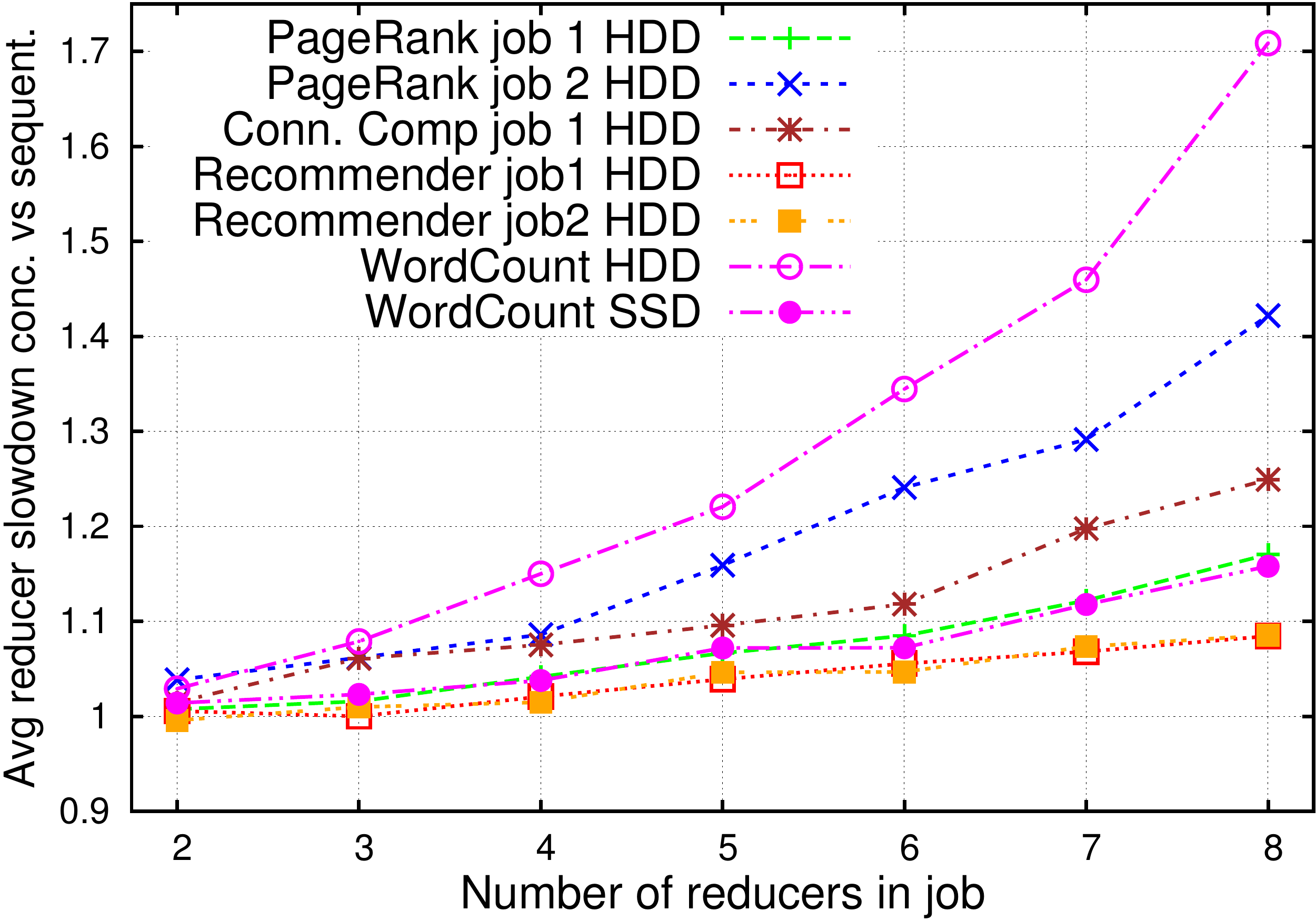}}
\caption{Memory elasticity for Spark, Tez and Flink jobs (a). Spilling vs paging (b). Impact of disk contention on Hadoop reducers (c).}
\end{figure*}

\subsection{Memory elasticity for Hadoop reducers}

Elasticity for reducers appears on their input side. Reducers need to read all
map outputs before starting the first call to the reduce function. Map outputs
are first buffered in memory. For a well-sized reducer this buffer never fills
up and it is never spilled. These in-memory map outputs are then merged
directly into the reduce functions. For an under-sized reducer the buffer fills
up while map outputs are being copied. In this case, the buffer is spilled and
reused.  The reduce function for an under-sized reducer is fed a merge of this
mixture of in-memory and on-disk data.

\vspace{0.01in} \noindent{\bf The impact of elasticity on reducer
runtime\hspace{0.2in}} \cref{fig:hadoop-elasticity} shows the memory
elasticity profiles for several Hadoop reducers.  In addition to the jobs in
\cref{fig:hadoop-elasticity} we also profiled the vector creation part of
HiBench's Bayesian Classification. Because this application has many jobs we
could not obtain the full elasticity profile for each individual job. Instead,
we inferred the maximum penalty for each job using the model described at the
end of this subsection. For the 8 distinct jobs we encountered, the maximum
penalties are: 1.8x, 1.67x, 2.65x, 1.42x, 3.32x, 1.37x, 1.75x and 1.42x. 

Two main insights arise from the results in \cref{fig:hadoop-elasticity}.
Most importantly, under-sized reducers incur only moderate penalties.  Given as
little as 10\% of ideal memory, 7 of the 10 reducers are between 1.1x and 1.85x
slower than ideal. Second, the penalties are comparable for a wide range of
under-sized memory allocations. For the WordCount reducer, for 83\%, 41\% and
10\% of ideal memory the penalties are: 1.84x, 1.83x and 1.82x.

\vspace{0.01in} \noindent{\bf Why the penalty varies among reducers\hspace{0.2in}} 
We found that the penalty correlates with the complexity of the reduce
function. More complex reducers are more CPU-intensive and thus are influenced less
by reading inputs from disk. A TeraSort reducer is very simple and shows one of
the highest penalties. On the other extreme, the NutchIndexing reducer is 
complex and shows almost no penalty.  To further analyze how reducer complexity
influences the penalty we added to the WordCount reduce function a number of
floating point operations (FPops) between two randomly generated numbers.
Adding 10, 50 and 100 FPops per reduce function call decreased the maximum
penalty from 2x to 1.87x, 1.65x and 1.46x.

We also found that the penalty correlates with the number of values
corresponding to each key processed by a reducer. A large number of values per
key leads to increased penalties because the read-ahead performed by the OS
becomes insufficient to bring all keys in memory and thus some reduce calls
will read on-disk data.  The Mahout recommender uses on average 1500 keys per
value for job1 and 15000 keys per value of job2. This explains their larger
penalty.

\vspace{0.01in} \noindent{\bf Why penalties are not larger\hspace{0.2in}}
The explanations for reducers are the same as the ones provided for mappers.

\vspace{0.01in} \noindent{\bf Modeling Hadoop reducer behavior\hspace{0.2in}}
\label{subsec:exec-infer}
An accurate model for a reducer can be obtained from two training runs, one
with an under-sized reducer and one with a well-sized reducer. Given these two
training runs, the model can infer the penalty for all other under-sized memory
allocations for that reducer. While the training runs are specific to an
application, the model we describe is generally applicable to any Hadoop
reducer.

Our model is based on three insights. First, the penalty incurred by spilling
to disk is extra penalty on top of the ideal runtime.  Second, the penalty is
proportional to the amount of spilled data. Third, the disk rate for reading
and writing spills remains constant for one reducer regardless of its memory
allocation. Thus, our model is centered around the equation: 

$T(notId) = T + spilledBytes(notId)/diskRate$

$T(notId)$ denotes the reducer runtime for an under-sized reducer allocated
$notId$ memory. $T$ is the runtime when that reducer is well-sized.
$spilledBytes(notId)$ is the amount of data spilled when being allocated
$notId$ memory. Finally, $diskRate$ is the average rate at which the
under-sized reducer uses the disk when reading and writing spills.

The two training runs provide $T(notId)$ and $T$. Next, $spilledBytes(notId)$
can be computed numerically from the reducer input size, the value of $notId$
and a few Hadoop configuration parameters. Thus, $diskRate$ can be obtained
from the equation. Once $T$ and $diskRate$ are known, any other $T(notId')$ can
be obtained numerically by computing the corresponding $spilledBytes(notId')$
value and plugging it into the equation.

\cref{fig:hadoop-reducer-model} shows the accuracy of our model for the
Hadoop reducers profiled in \cref{fig:hadoop-elasticity}. The value of
$notId$ chosen for the under-sized training run was 52\% of optimal. Any other
under-sized amount would have sufficed. The accuracy of the model is within
5\% for most cases.

\vspace{0.01in} \noindent{\bf The shape of the memory elasticity profile\hspace{0.2in}}
We now explain the sawtooth-like shape of the memory elasticity profiles from
\cref{fig:hadoop-elasticity}. The key insight is that the penalty is
proportional to the amount of spilled data.

The peaks of the sawtooth are caused by a reducer spilling nearly all its input
data to disk. This behavior appears for several under-sized allocations.
For example, assume a reducer with a 2GB shuffle buffer and 2.01GB of input
data. When the buffer is filled, all 2GB are spilled to disk.  With a 500MB
shuffle buffer, the reducer with 2.01GB input would spill 500MB four times,
again for a total of 2GB spilled data.

There are several cases in which decreasing the memory allocation also
decreases the penalty (e.g., WordCount with 52\% vs 83\% of ideal memory).
This is caused by a decrease in the amount of spilled data. Given a 2GB shuffle
buffer and a 2.01GB input size, a reducer spills 2GB to disk but given a 1.5GB
shuffle buffer it spills only 1.5GB to disk and keeps 510MB in memory. 

One may argue that the static threshold used by Hadoop for spilling is
inefficient and that Hadoop should strive to spill as little as possible. In
this argument, the reducer with 2.01GB input and a 2GB shuffle buffer would
spill 10MB only. Such a proposal actually strengthens the case for memory
elasticity as the penalties decrease (due to less spilled data) and can be
modeled similarly.

\subsection{Elasticity for Spark, Tez and Flink}
\cref{fig:sparktez-elasticity} shows that memory elasticity also applies
to Spark. For Spark we profiled a task performing a sortByKey operation
(TeraSort) and one performing a reduceByKey operation (WordCount). Internally,
Spark treats the two cases differently. A buffer is used to store input data
for sortByKey and a hashmap for reduceByKey. Both data structures are spilled
to disk when a threshold is reached. Despite the differences both tasks show
elasticity.

\cref{fig:sparktez-elasticity} shows that the elasticity profile for
Spark resembles that of Hadoop reducers. Given the similarities we were able to
extend our Hadoop reducer model to Spark sortByKey tasks (TeraSort). The
difference between the Hadoop and Spark TeraSort model is that for Spark we
also learn an expansion factor from the under-sized training run. This is
because Spark de-serializes data added when adding it to the shuffle buffers.
\cref{fig:hadoop-reducer-model} shows that the accuracy of the Spark
model is well within 10\%, matching that of Hadoop's model.

~\cref{fig:sparktez-elasticity} also shows the memory elasticity profiles for
two Tez reducers. The elasticity profile for Tez is similar to those for Spark
and Hadoop. We extended our Hadoop reducer model to Tez reducers by accounting
for the fact that in Tez, map outputs stored on the same node as the reducer
are not added to shuffle memory but are instead read directly from disk.
\cref{fig:hadoop-reducer-model} shows that the accuracy of our Tez model is
equally good.

Finally, \cref{fig:sparktez-elasticity} also shows one Flink reducer.
Flink stands out with its low penalty between 70\% and 99\% of optimal memory
which suggests a different model is needed. We plan to purse this as part of
our future work.

\subsection{Spilling vs paging}
\label{subsec:spilling_vs_paging}
Why do frameworks implement spilling mechanisms and do not rely on
tried-and-tested OS paging mechanisms for under-sized tasks? To answer, we
provisioned Hadoop with enough memory to avoid spilling but configured cgroups
such that part of the memory is available by paging to a swapfile.
\cref{fig:paging_vs_spilling} shows the results for the Hadoop Wordcount
reducer. Paging wins when a task gets allocated close to ideal memory (0.7 or
more on the x-axis) because it only writes to disk the minimum necessary while
Hadoop spills more than necessary. However, spilling beats paging for smaller
memory allocations because the task's access pattern does not match the LRU order
used by paging. \cref{fig:paging_vs_spilling} also shows that paging
greatly increases garbage collection (GC) times because the GC touches memory
pages in a paging-oblivious manner. We also see that the SSD significantly
outperforms the HDD due to more efficient page-sized (4k) reads and writes.
Using 2MB transparent huge pages (THP) did not improve results for either the
SSD or HDD since THP is meant to alleviate TLB bottlenecks not improve IO
throughput.

\subsection{Memory elasticity and disk contention}
\label{subsec:contention}

Since memory elasticity leverages secondary storage, it is interesting to
understand the impact of disk contention when several under-sized tasks are
collocated.

The impact of disk contention depends on how well provisioned the local storage
is on nodes relative to compute. The ratio of cores to disks can give a sense
of how many under-sized tasks can compete, in the worst case, for the same disk
(a task usually requires at least one core). In current data centers the ratio
is low. In~\cite{oust-nsdi15}, the authors mention ratios between 4:3 and 1:3
for a Facebook 2010 cluster. Public provider offerings also have low core to
disk ratios. The list of high-end Nutanix hardware
platforms~\cite{nutanix-specs} shows plenty of offerings with a ratio of less
than 2.5:1 and as low as 0.66:1. Nutanix has more than two thousand small and
medium size clusters at various enterprises~\cite{nutanix-socc16}. 

Nevertheless, not all clusters are equally well provisioned. Thus, we analyzed
the degree to which memory elasticity can produce disk contention by varying
the number of under-sized Hadoop reducers that spill concurrently to the same
disk. We start between 2 and 8 under-sized reducers each within 1 second of the
previous. This is akin to analyzing disk contention on nodes with a core to
disk ratio ranging from 2:1 to 8:1. We focused on reducers because they spill
more data than the mappers (GBs is common).

We measured the slowdown in average reducer runtime when all reducers run
concurrently compared to the case where they run sequentially.
\cref{fig:concurrency} shows the results. Several reducers (PageRank
job1, Recommender job1,2) show minimal slowdown (at most 15\% degradation
for 8 concurrently under-sized reducers). In the other extreme, running 8
under-sized WordCount reducers concurrently leads to a 70\% degradation when an
HDD is used but that is reduced to just 15\% when moving the spills to SSD. In
conclusion, disk contention is a manageable potential side effect but should
nevertheless be taken into consideration when leveraging elasticity.

\subsection{Final considerations}
\label{sec:study:final}

\noindent{\bf Does elasticity cause increased GC?\hspace{0.2in}}
For Hadoop and Tez reducers, GC times remain constant when tasks are
under-sized. For Hadoop mappers, GC times slowly increase as the memory
allocation decreases but they remain small in all cases. Overall, Hadoop does a
good job of mitigating GC overheads by keeping data serialized as much as
possible. For Spark, GC times increase sub-linearly with an increase in task
runtime. Interestingly, GC times are a larger portion of task runtime for
well-sized tasks because spill episodes limit the amount of data that needs to
be analyzed for GC.

\noindent{\bf Feasibility of modeling\hspace{0.2in}}
Our models for Hadoop, Tez and Spark are based on two training runs, one
under-sized and one well-sized. Related work shows that a large fraction of
jobs in current data centers are recurring, have predictable resource
requirements and compute on similar
data~\cite{jockey,netw-aware-sched-sig15,s.agarwal-nsdi2012}. Thus, instead of
training runs, one can use prior runs of recurring jobs. Alternatively, if no
prior runs are available, the two training runs can be performed efficiently on
a sample of the job's input data. 

\noindent{\bf Sensitivity to configuration changes\hspace{0.2in}}
We repeated our experiments on two different hardware platforms (Dual Intel Xeon
E5-2630v3 + 40Gb NIC, Dual Opteron 6212 + 10GB NIC), two OSes (RHEL 7, Ubuntu
14.04), three different disk configurations (HDD, SDD, 2*HDD in RAID 0), three
IO schedulers (CFS, deadline, noop) and three JVMs (HotSpot 1.7, 1.8, OpenJDK
1.7). The changes did not impact the memory elasticity profiles or the
accuracy of our model.

\section{Applying elasticity to cluster scheduling. Case study: Apache YARN} 
\label{sec:yarn-me}

In this section, we explore the benefits that memory elasticity can provide in
cluster scheduling by integrating memory elasticity into Apache
YARN~\cite{yarn-socc13}. We chose YARN because it is very popular and provides
a common resource management layer for all popular frameworks tested in
\S\ref{sec:study}. Moreover, several recent research efforts from large
Internet companies were validated with implementations on top of YARN
~\cite{rayon, morpheus, carbyne, graphene}. In addition, we also discuss how
the elasticity principles can be adopted to the Mesos \cite{mesos} resource
manager.

Scheduling using memory elasticity is an NP-hard problem because it contains as
a special case NP-hard variants of the RCPSP problem~\cite{rcpsp}, a well-known
problem in operations research. Nevertheless, we show that the benefits of memory
elasticity can be unveiled even using a simple approach.

\begin{figure*}[t]
	\vspace{-0.2in}
	\subfloat[All tasks \textit{regularly} allocated.]
	{\label{fig:benefits:regular}\includegraphics[width=.5\textwidth]{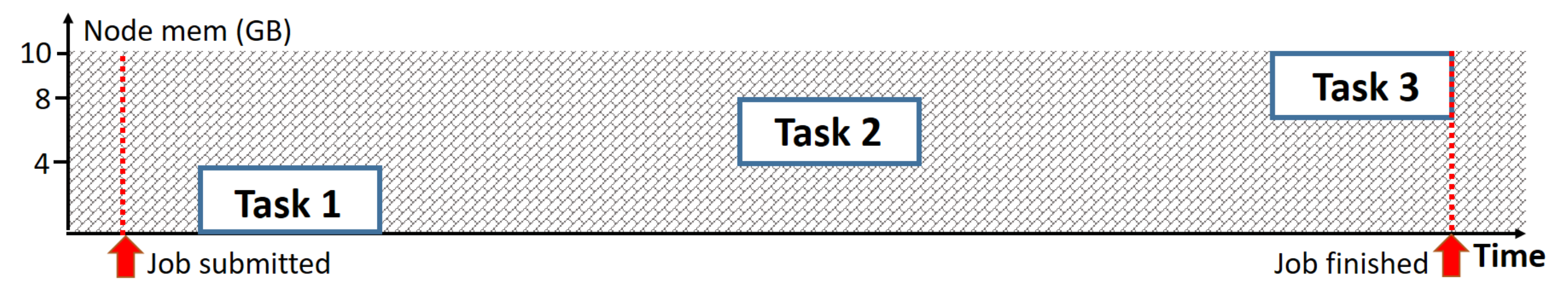}}
	\hfill
	\subfloat[All tasks \textit{elastically} allocated.]
	{\label{fig:benefits:elastic}\includegraphics[width=.5\textwidth]{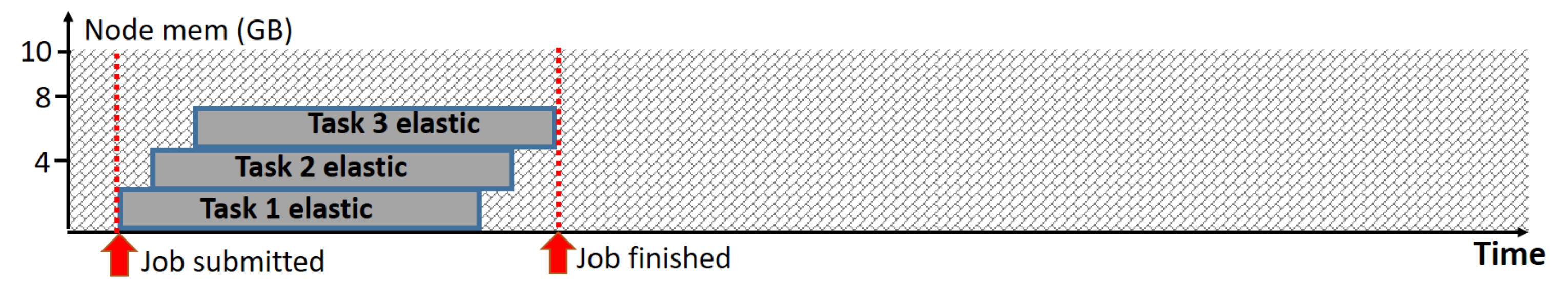}}
	\hfill
	\caption{Example of improvement in job completion time for a simple 3-task job and one highly utilized node.}\label{fig:benefits}
\end{figure*}

\subsection{Overview} \label{sec:yarn-me:overview}

YARN distributes cluster resources to the jobs submitted for execution.  A
typical job may contain multiple tasks with specific resource requests.  In
YARN, each task is assigned to a single node, and multiple tasks may run
concurrently on each node, depending on resource availability. The scheduler
has a global view of resources and queues incoming jobs according to cluster
policy (e.g., fair sharing with respect to resource usage).

\vspace{0.01in} \noindent{\bf Notation} We use the term {\it regular} to refer
to the memory allocation and runtime of well-sized tasks and the term {\it
elastic} for under-sized tasks. We further refer to our elasticity-aware
implementation as YARN-ME.

\vspace{0.01in} \noindent{\bf Benefits} As previously discussed, memory
elasticity trades-off \textit{task execution time} for \textit{task memory
allocation}. When applied to cluster scheduling it becomes a trade-off between
\textit{task queuing time} and \textit{task completion time}.  A task normally
has to wait until enough memory becomes available for it but executing it with
less memory may reduce or eliminate its waiting time.  Since the completion
time of a task is the sum of waiting time plus runtime, a significant decrease
in waiting time may outweigh an increase in runtime due to elasticity and
overall lead to faster task completion times. YARN-ME turns this task level
trade-off into improved job completion time and improved cluster-wide memory
utilization.  

\cref{fig:benefits} illustrates how memory elasticity benefits scheduling using
a simple example of a 3-task job scheduled on a single, highly utilized node.
\cref{fig:benefits:regular} presents a timeline of task execution for vanilla
YARN. Tasks 2 and 3 incur queuing times much larger than their execution times.
In \cref{fig:benefits:elastic}, using memory elasticity, the scheduler launches
all tasks soon after job submission, resulting in the job completing in less
than 30\% of its original time, despite its tasks now taking twice as long to execute.

\subsection{System design} \label{sec:solution:model}
Two main additions are needed to leverage memory elasticity in YARN.

\vspace{0.01in} \noindent{\bf Metadata regarding task memory elasticity}
Reasoning about memory elasticity at the scheduler level requires additional
knowledge about the submitted tasks. The scheduler needs to know the 
regular execution time of a task (ideal duration), and the minimal
amount of memory for a regular allocation (ideal memory). It also
needs to understand the performance penalty when receiving less than
its ideal memory amount. The metadata are obtained using the profiling
and modeling techniques described in \S\ref{sec:study:final}.

\vspace{0.01in} \noindent{\textbf{The timeline generator}} YARN-ME uses a
timeline generator to provide an estimate of a job's evolution (the completion
times of its tasks and of the whole job). In doing this, it accounts for the
expected memory availability in the cluster. The generator simply iterates over
all the nodes, adding up the task duration estimates of the executing and
queued tasks. In effect, the generator builds simple timelines for each node,
which it then merges to obtain information about each job's timeline. The
generator runs periodically, every heartbeat interval, since during such a
period all healthy nodes report their status changes. It also runs when a new
job arrives or an existing one is prematurely canceled.

\subsection{Scheduler decision process} \label{sec:solution:decision}

\vspace{0.01in} \noindent{\bf Main heuristic} YARN-ME aims to reduce job
completion time by leveraging memory elasticity. As such, an elastic task
cannot be allowed to become a straggler for its respective job.  Therefore,
{\it an elastic allocation is made for a task that cannot be scheduled regularly
\underline{iff} its expected completion time does not exceed the current
estimated completion time of its job}. 

\vspace{0.01in} \noindent{\bf Disk contention awareness} As shown
in~\S\ref{subsec:contention} scheduling too many elastic tasks concurrently on a
node may lead to disk contention. YARN-ME incorporates disk contention
awareness. As shown in~\S\ref{subsec:contention}, obtaining the task metadata
involves computing the amount of disk bandwidth required by an elastic task.
YARN-ME uses this information as well as a disk bandwidth budget reserved per
node for the use of elastic tasks. YARN-ME conservatively prohibits new elastic
tasks from being scheduled on nodes where the elasticity disk budget would be
exceeded.

\vspace{0.01in} \noindent{\bf Node reservations} In YARN, if a node has
insufficient resources to satisfy the job at the head of the queue, no
allocation is performed, and that job reserves the node. As long as
the reserving job still has pending tasks, no other job would be able to
schedule tasks on the reserved node. This helps mitigate resource starvation by
ensuring that jobs with large memory requirements also get the chance to
schedule. To account for this, we adjusted the \textit{timeline generator} to
take reservations into account when building its estimates. Additionally,
YARN-ME allows tasks of other jobs to be allocated on a reserved node, but only
if this does not hinder tasks of the reserved job.

\vspace{0.01in} \noindent{\bf Additional constraints} Schedulers may set
additional constraints for their jobs, such as running on a data-local node
only, or forcing certain tasks to start only after others have completed. Our
design is orthogonal to such constrains, requiring only the tweaking of the
\textit{timeline generator} to function. 

\Cref{alg:decision} presents the decision process of YARN-ME.  Lines 8-11 apply
the main heuristic underlying YARN-ME. Line 7 includes the check for disk
contention. Additionally, lines 7 and 10 always consider the minimum amount of
memory that yields the lowest possible execution time, leveraging the
behavior of elasticity described in \S\ref{sec:study}.

\setlength{\textfloatsep}{0pt}
\begin{algorithm}[t!]
	\caption{YARN-ME decision process pseudocode.}
	\label{alg:decision}
	\begin{algorithmic}[1]
		\While{\textsc{Job Queue} is not empty}
			\State \textbf{J} $\leftarrow$ next job in \textsc{Job Queue}
			\ForAll{node \textbf{N} in \textsc{Nodes}}
				\State \textbf{T} $\leftarrow$ next task of (\textbf{N}'s reserved job \textbf{or J})
				\If{\textbf{T} \textit{regularly} fits on \textbf{N}}
					\State allocate \textbf{T} on \textbf{N}, \textbf{regular}
				\ElsIf{\textbf{T} \textit{elastically} fits on \textbf{N}}
					\State get \textsc{Timeline Generator} info for \textbf{J}
					\If{\textbf{T} \textit{elastically} finishes before \textbf{J}}
						\State allocate \textbf{T} on \textbf{N}, \textbf{elastic}
					\Else
						\State do not schedule anything new on \textbf{N}
						\State \textbf{reserve N} for \textbf{J}, if not already reserved
					\EndIf
				\EndIf
				\If{\textbf{T} was allocated}
					\State \textbf{unreserve N} if reserved
					\State resort the \textsc{Job Queue}
					\State \textbf{J} $\leftarrow$ next job in \textsc{Job Queue}
				\EndIf
			\EndFor
		\EndWhile
	\end{algorithmic}
\end{algorithm}

\section{Discussion: Mesos} \label{sec:discussmesos}

Other schedulers beyond YARN can also be extended to use memory elasticity.  We
next review the main differences between Mesos~\cite{mesos} and YARN and argue
that they do not preclude leveraging memory elasticity in Mesos.

\vspace{0.01in} \noindent{\bf Queuing policy} Mesos uses Dominant Resource
Fairness (DRF)~\cite{drf}, a multi-resource policy, to ensure fairness. Thus,
the job queue may be sorted differently compared to YARN's policies. This does
not restrict memory elasticity as it only dictates which job to serve next.

\vspace{0.01in} \noindent{\bf Decision process} Mesos decouples scheduling
decisions from node heartbeats. Thus, a job may be offered resources from
several nodes at the same time. This does not restrict memory elasticity since
the job needs to consider each node from the offer separately (a task can only
run on one node), so memory elasticity can be applied for every node in the
offer.

\vspace{0.01in} \noindent{\bf Global vs local decisions} Mesos gives jobs the
ability to accept or reject resource offers while YARN decides itself what each
job receives. Thus, in Mesos, jobs can decide individually whether to use
elasticity or not. If a decision based on global cluster information (like in
YARN) is desired, jobs can express constraints (locality, machine
configuration) with Mesos filters that can be evaluated by Mesos before making
resource offers.

\section{Cluster experiments}
\label{sec:realevaluation}

\begin{table*}[t]
	\begin{center}    
		\scalebox{0.80}{
			\begin{tabular}{|c|c|c|c|c|c|c|c|c|c|c|c|c|c|}
				\hline
				\multirow{3}{*}{Application} & \multirow{3}{0.5cm}{\centering \# jobs} & \multirow{3}{0.75cm}{\centering Input GB} & \multirow{3}{0.75cm}{\centering \# maps} & \multirow{3}{1.1cm}{\centering \# reduces} & \multicolumn{4}{c|}{Penalties} & \multicolumn{4}{c|}{Memory GB} & \multirow{3}{1.7cm}{\centering Inter-arrival (IA) time} \\
				\cline{6-13}
				& & & & & \multicolumn{2}{c|}{\nth{1} job} & \multicolumn{2}{c|}{\nth{2} job} & \multicolumn{2}{c|}{\nth{1} job} & \multicolumn{2}{c|}{\nth{2} job} & \\
				\cline{6-13}
				& & & & & map & reduce & map & reduce & map & reduce & map & reduce & \\
				\hline
				Pagerank & 2 & 550 & 1381 / 1925 & 275 & 1.3 & 1.22 & 1.25 & 1.75 & 1.7 & 3.7 & 1.5 & 6.8 & 120s \\
				\hline
				WordCount & 1 & 540 & 2130 & 75 & 1.35 & 1.9 & - & - & 1.7 & 5.4 & - & - & 30s \\
				\hline
				Recommender & 2 & \multirow{1}{*}{250} & 505 / 505 & \multirow{1}{*}{100} & \multirow{1}{*}{1.3} & \multirow{1}{*}{2.6} & \multirow{1}{*}{1.3} & \multirow{1}{*}{3.3} & \multirow{1}{*}{2.4} & \multirow{1}{*}{2.8} & \multirow{1}{*}{2.4} & \multirow{1}{*}{3.8} & \multirow{1}{*}{120s} \\
				\hline     
			\end{tabular}  
		}  
	\end{center}    
	\vspace{-0.1in}
	\caption{Characteristics of the evaluated applications.}
	\label{tab:real-desc}
	\vspace{-0.1in}
\end{table*}   

\begin{figure*}[ht] 
	\vspace{-0.1in}
	\subfloat[Cluster memory utilization -- 5 runs]
	{\label{fig:real-util-pagerank5j}\includegraphics[width=.33\textwidth]{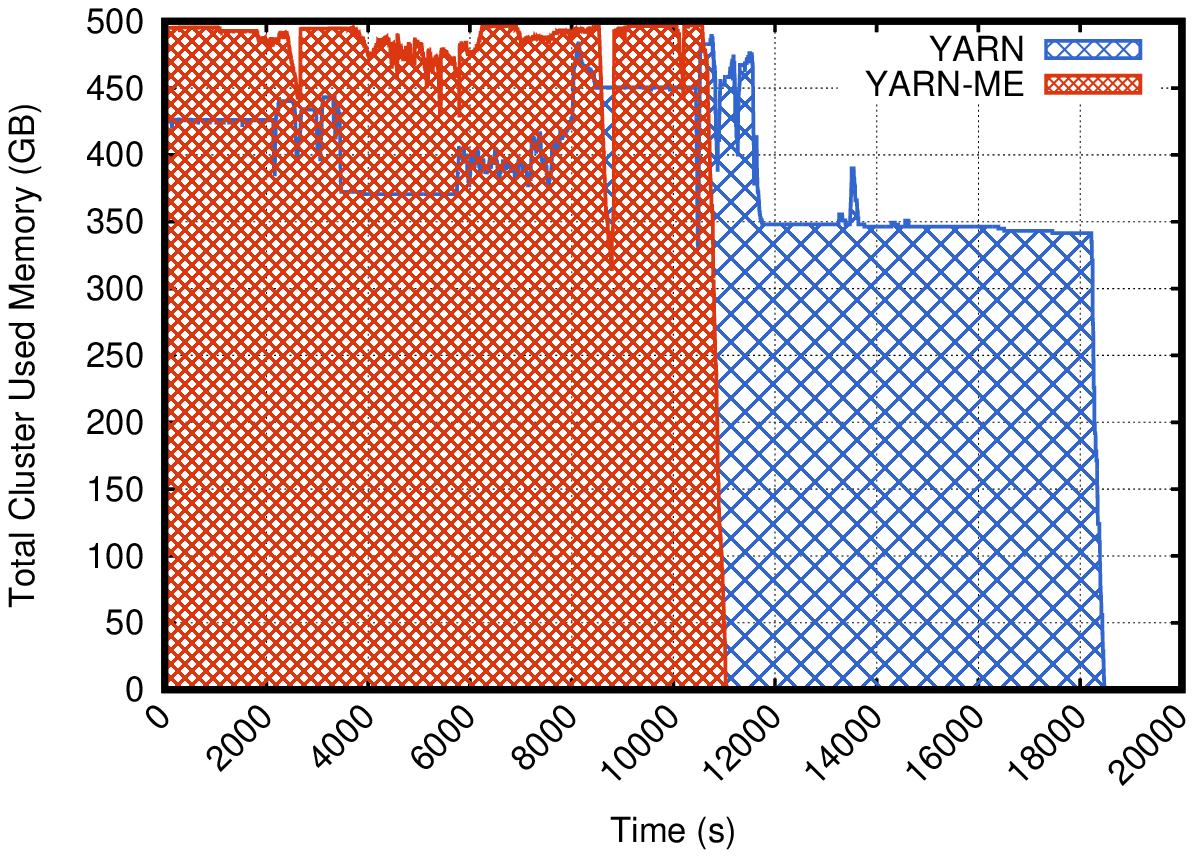}}
	\hfill
	\subfloat[\textit{elastic} tasks -- 5 runs]
	{\label{fig:real-ela-cont-pagerank5j}\includegraphics[width=.33\textwidth]{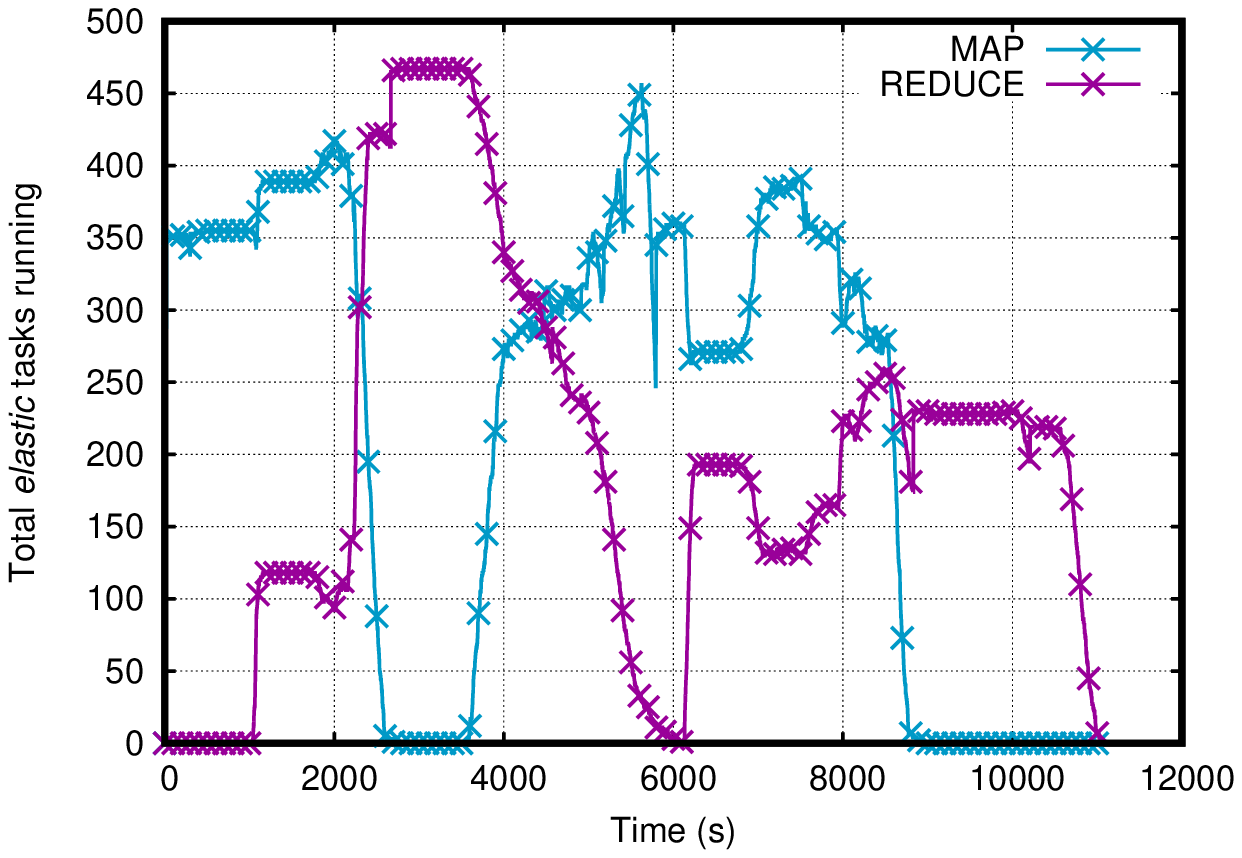}}
	\hfill
	\subfloat[Improvement over YARN]
	{\label{fig:real-jrt-pagerank}\includegraphics[width=.33\textwidth]{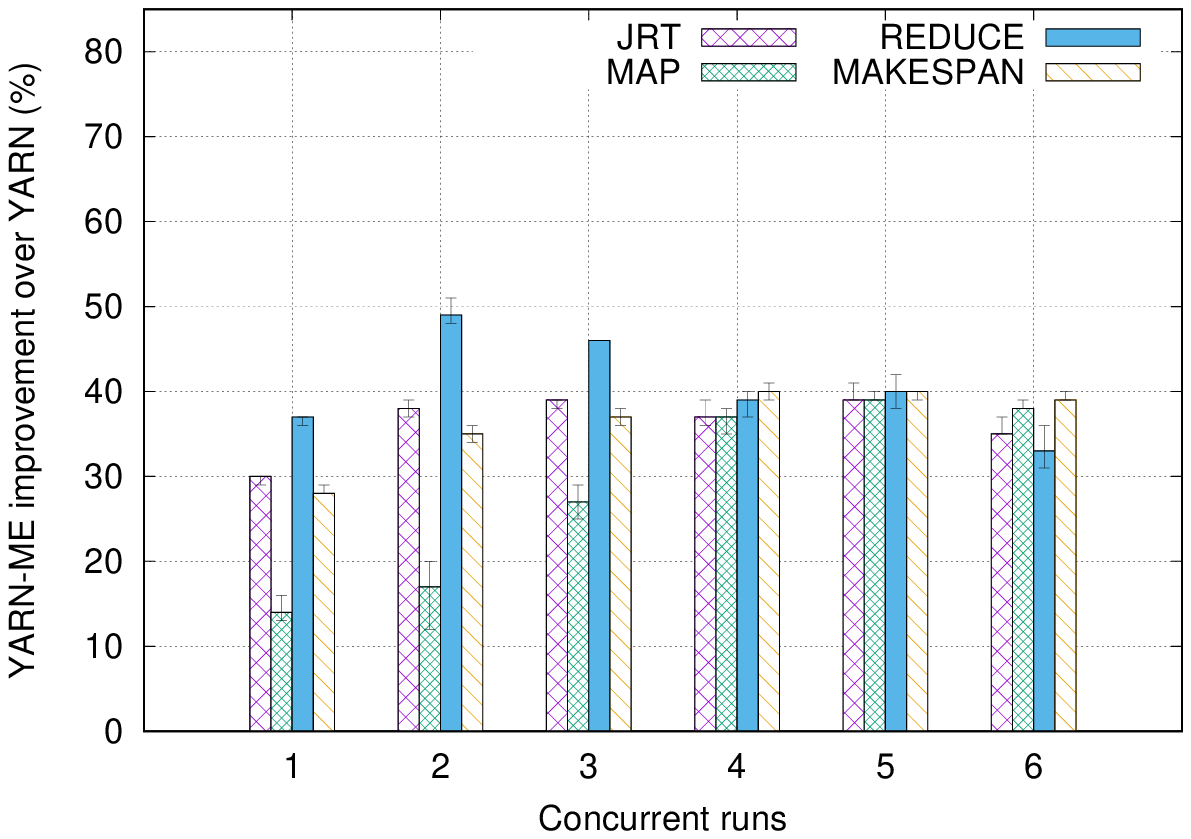}}
	\caption{YARN-ME vs YARN for running Pagerank on 50 nodes. \cref{fig:real-util-pagerank5j} shows the timeline of cluster memory utilization. \cref{fig:real-ela-cont-pagerank5j} shows the timeline of tasks scheduled \textit{elastically}.
	\cref{fig:real-jrt-pagerank} reports improvement w.r.t. \textit{average job runtime} (JRT), \textit{average job phase time} (map, reduce), and makespan. }
\end{figure*}

\begin{figure*}[t!] 
	\vspace{-0.2in}
	\hfill
	\subfloat[WordCount]
	{\label{fig:real-jrt-wordcount}\includegraphics[width=.33\textwidth]{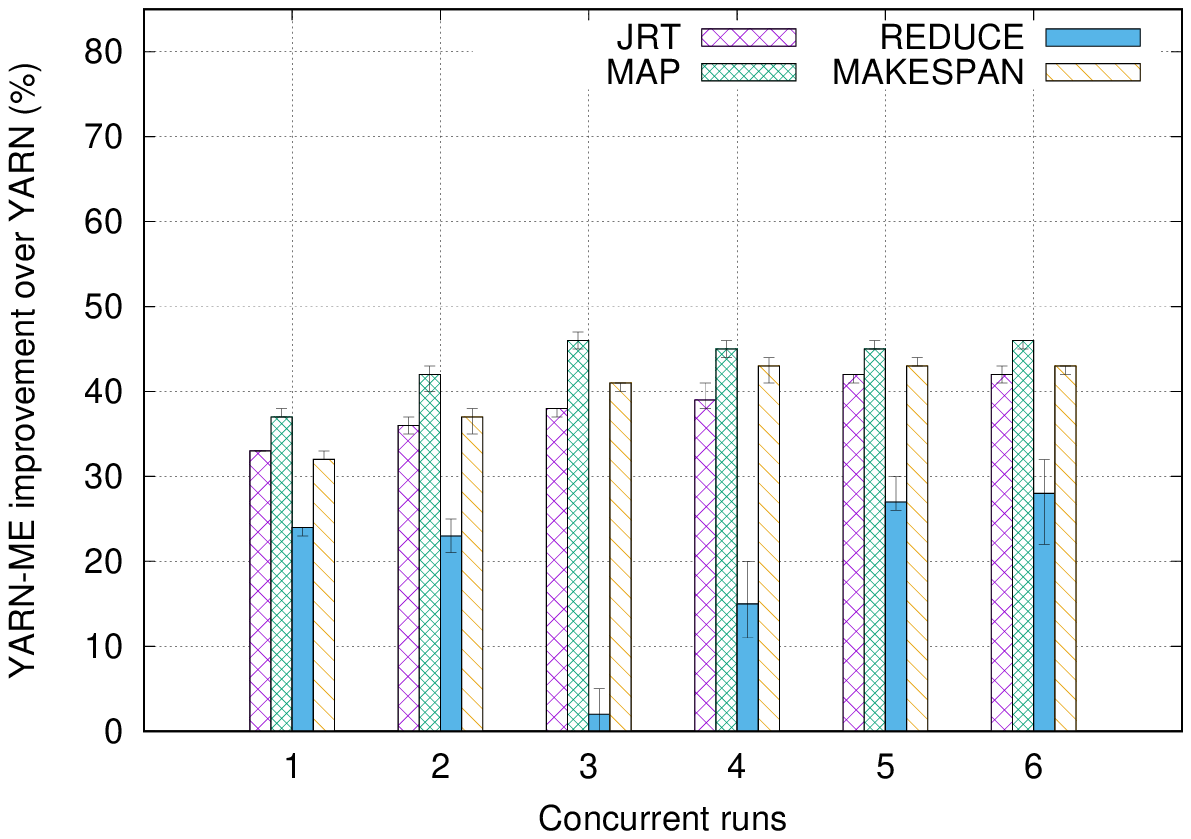}}
	\hfill
	\subfloat[Recommender]
	{\label{fig:real-jrt-mahoutir}\includegraphics[width=.33\textwidth]{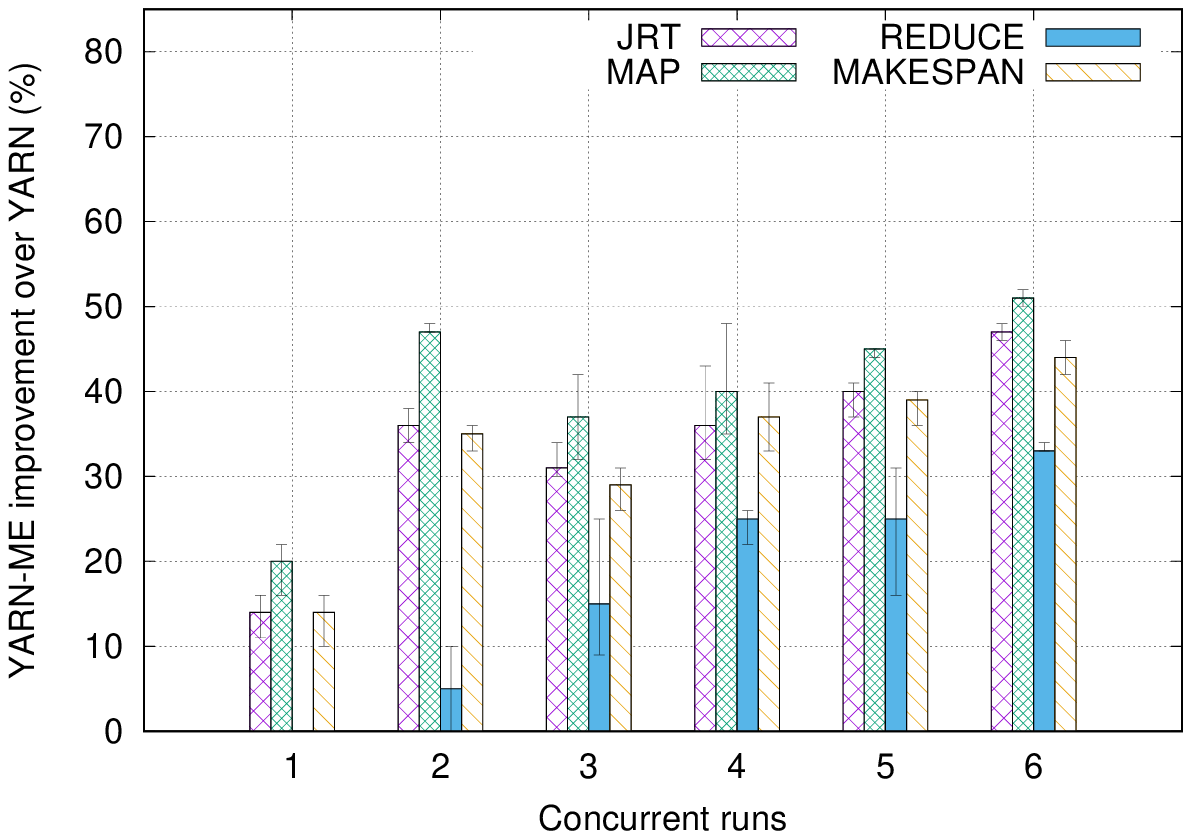}}
	\subfloat[Mixed job trace -- 14 jobs]
	{\label{fig:real-jrt-trace}\includegraphics[width=.33\textwidth]{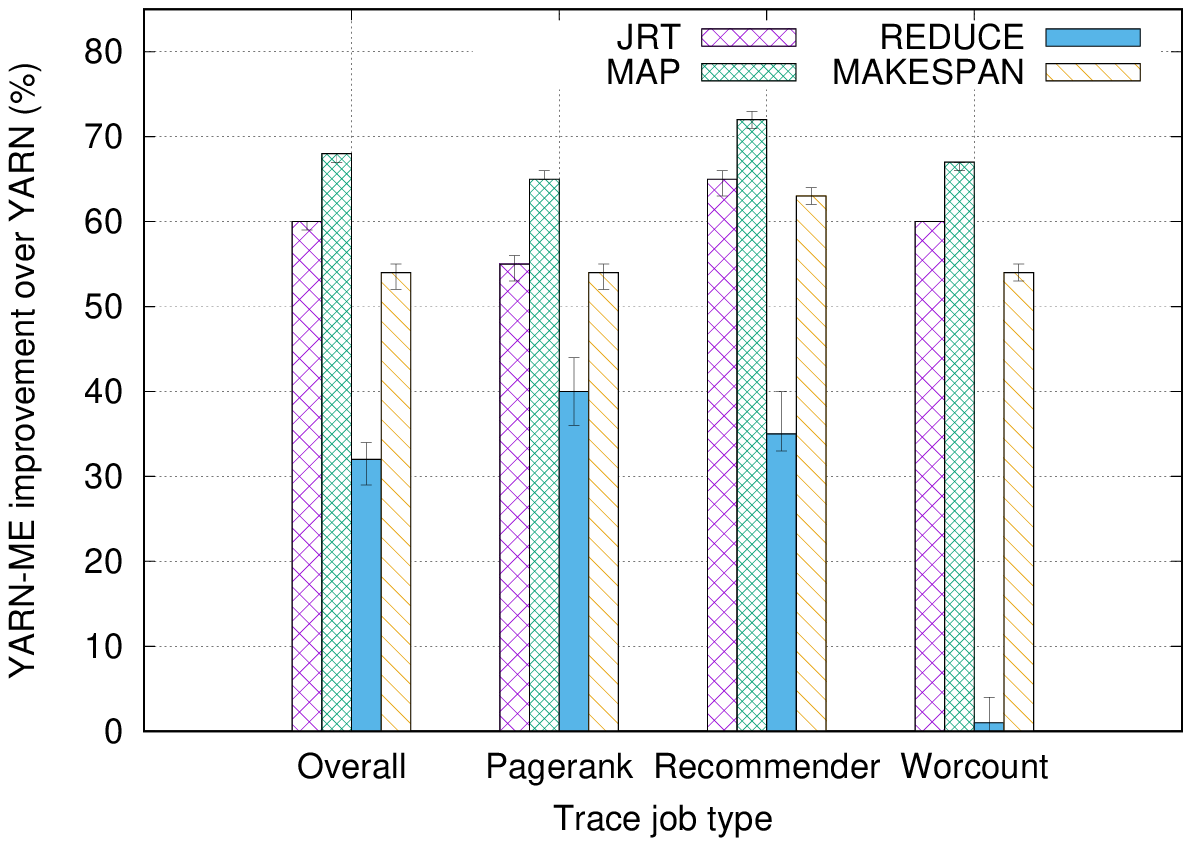}}
\caption{Further experiments on 50 nodes. Improvement of YARN-ME over YARN w.r.t. \textit{average job runtime} (JRT), \textit{average job phase runtime} (map, reduce), and makespan. We report average, min. and max. over 3 iterations. \cref{fig:real-jrt-trace} reports results for a mixed trace of jobs: 3x Pagerank, 3x Recommender, 8x Wordcount.}
	\vspace{-0.2in}
\end{figure*}

We next showcase the benefits of memory elasticity by comparing YARN-ME to YARN.

\subsection{Methodology}

\noindent{\bf Setup \hspace{0.2in}} We use a 51-node cluster (50 workers and 1
master), limiting the scheduler to 14 cores per node (out of 16 cores we
reserve 2 for the YARN NodeManager and for the OS) and 10GB of RAM. The exact
amount of RAM chosen is not important (we could have chosen any other value),
what is important is the ratio of ideal task memory requirements to node
memory. Each node has one 2 TB SATA HDD.  YARN-ME was implemented on top of
Apache YARN 2.6.3~\cite{yarn-socc13}. Disk spills use Direct I/O so that the OS
buffer cache does not mask performance penalties due to elasticity.

We ran WordCount, PageRank and Mahout Item Recommender Hadoop applications. We
chose them because they represent small, medium and large penalties encountered
for reducers in \S\ref{sec:study} (mapper penalties span a much smaller
range than reducers and are lower). We configured the jobs as described in
\cref{tab:real-desc}. We executed each type of application separately
(homogeneous workload) and all applications simultaneously (heterogeneous
workload). For the homogeneous workloads, we varied the number of concurrent
runs for each type of application.  The start of each run is offset by the
inter-arrival (IA) time mentioned. The IA time is chosen proportionally to
application duration such that map and reduce phases from different jobs can
overlap.

For each application we first perform one training run using ideal memory to
obtain the ideal task runtime. We multiply this by the penalties measured in
\S\ref{sec:study} to obtain task runtimes for various under-sized
allocations.

\noindent{\bf Metrics\hspace{0.2in}} We compare average job runtime, trace
makespan and average duration of \textit{map} and \textit{reduce} phases. By
job runtime we mean the time between job submission and the end of the
last task in the job. Similarly, a map or reduce phase represents the time
elapsed between the first request to launch such a task and the finish time of
the last task in the phase. Each experiment is run for 3 iterations, and we
report the average, minimum and maximum values.

\subsection{Experiments}

\noindent{\bf Benefits for memory utilization \hspace{0.2in}}
\cref{fig:real-util-pagerank5j} shows the benefits of using memory elasticity
on both cluster utilization and makespan, for an execution of 5 Pagerank runs.
YARN-ME successfully makes use of idle memory, bringing total memory
utilization from 77\% to 95\%, on average, and achieving a 39\% reduction in
makespan.  We also observe, by comparing the results in
\cref{fig:real-jrt-pagerank,fig:real-util-pagerank5j} that the gain in job
runtime is proportionally much higher than that of memory reclaimed.
\cref{fig:real-ela-cont-pagerank5j} shows how YARN-ME assigns the memory slack
to tasks.

\noindent{\bf Benefits for homogeneous workloads\hspace{0.2in}} We next
show that YARN-ME can provide benefits for the jobs in \cref{tab:real-desc}.
\cref{fig:real-jrt-pagerank,fig:real-jrt-mahoutir,fig:real-jrt-wordcount} show
the improvement of YARN-ME compared to YARN vs. the number of runs. YARN-ME's
benefits hold for all jobs.

We find that the Recommender, which has the highest penalties we have observed
for reducers, achieves up to 48\% improvement. We also find that mappers always
benefit noticeably from elasticity, a direct consequence of their modest
penalties. Pagerank's lower-penalty reducers yield an improvement of 30\% even
for a single concurrent run, peaking at 39\% with 5 runs.  Wordcount achieves a
peak improvement of 41\%, despite reducer gains being lower, due to higher
penalties. The reduction in average job runtime steadily increases across runs.
For 3 concurrent Wordcount runs, the number of reducers leads only one out of
the 3 jobs to be improved, but the map phase still reaches improvements of
46\%.

\noindent{\bf Benefits for heterogeneous workloads\hspace{0.2in}} YARN-ME
achieves considerable gains even under a heterogeneous workload composed of all
the jobs from \cref{tab:real-desc}. We start 5 jobs at the same time (1
Pagerank, 1 Recommender and 3 Wordcount) and then submit a new job every 5
min., until we reach a total of 14 jobs (3 Pagerank, 3 Recommender and 8
Wordcount). Each job is configured according to \cref{tab:real-desc}.
\cref{fig:real-jrt-trace} shows overall improvement and breakdown by job type.
YARN-ME improves average job runtime by 60\% compared to YARN. The \textit{map}
phase duration is reduced by 67\% on average overall, and by up to 72\% for all
Recommender jobs.

\section{Simulation experiments}
\label{sec:simevaluation}

We use simulations to evaluate YARN-ME's benefits and its robustness to
mis-estimations on a much wider set of workload configurations than we can run
in the real cluster.

\begin{figure*}[t]
	\subfloat[YARN-ME vs YARN (avg job runtime)  \label{fig:ELASTIC_vs_YARN_jrt}]
	{\includegraphics[width=.23\textwidth,angle=270]{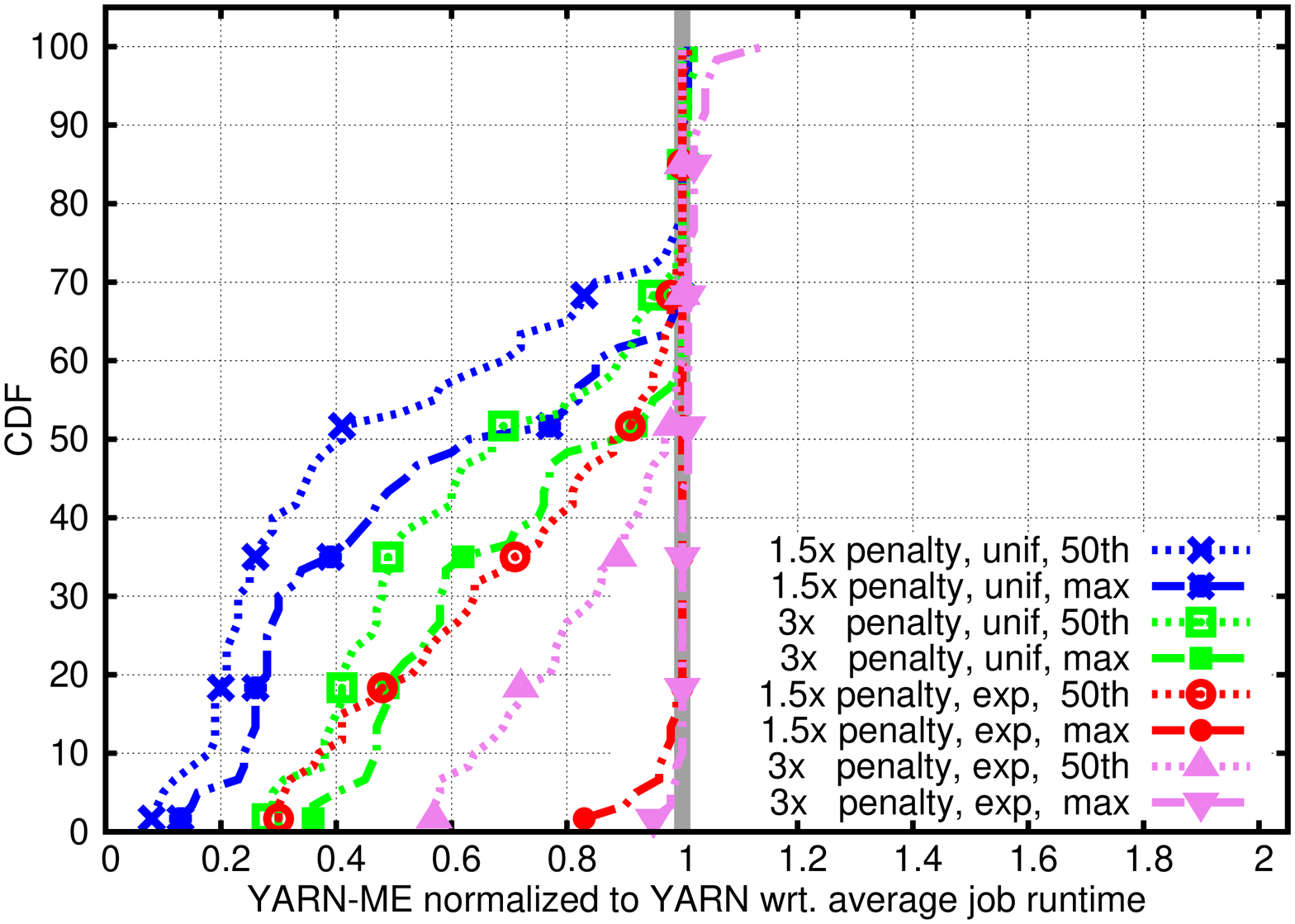}}
	\hfill
	\subfloat[Impact of scaling trace size\label{fig:ELASTIC_vs_YARN_largetrace}]
	{\includegraphics[width=.23\textwidth,angle=270]{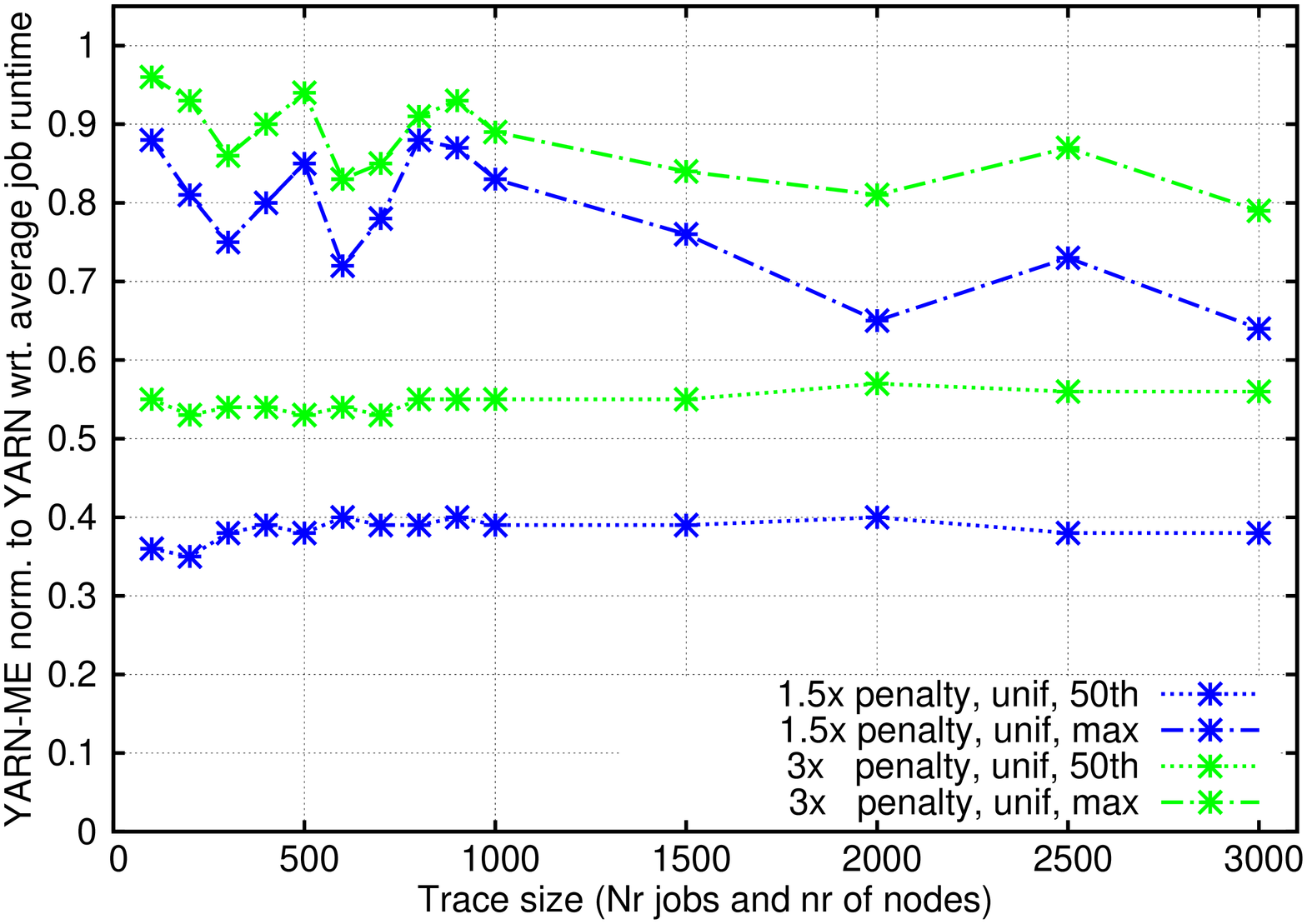}}
	\hfill
	\subfloat[YARN-ME vs Meganode (avg job runtime) \label{fig:ELASTIC_vs_meganode_jrt}]
	{\includegraphics[width=.23\textwidth,angle=270]{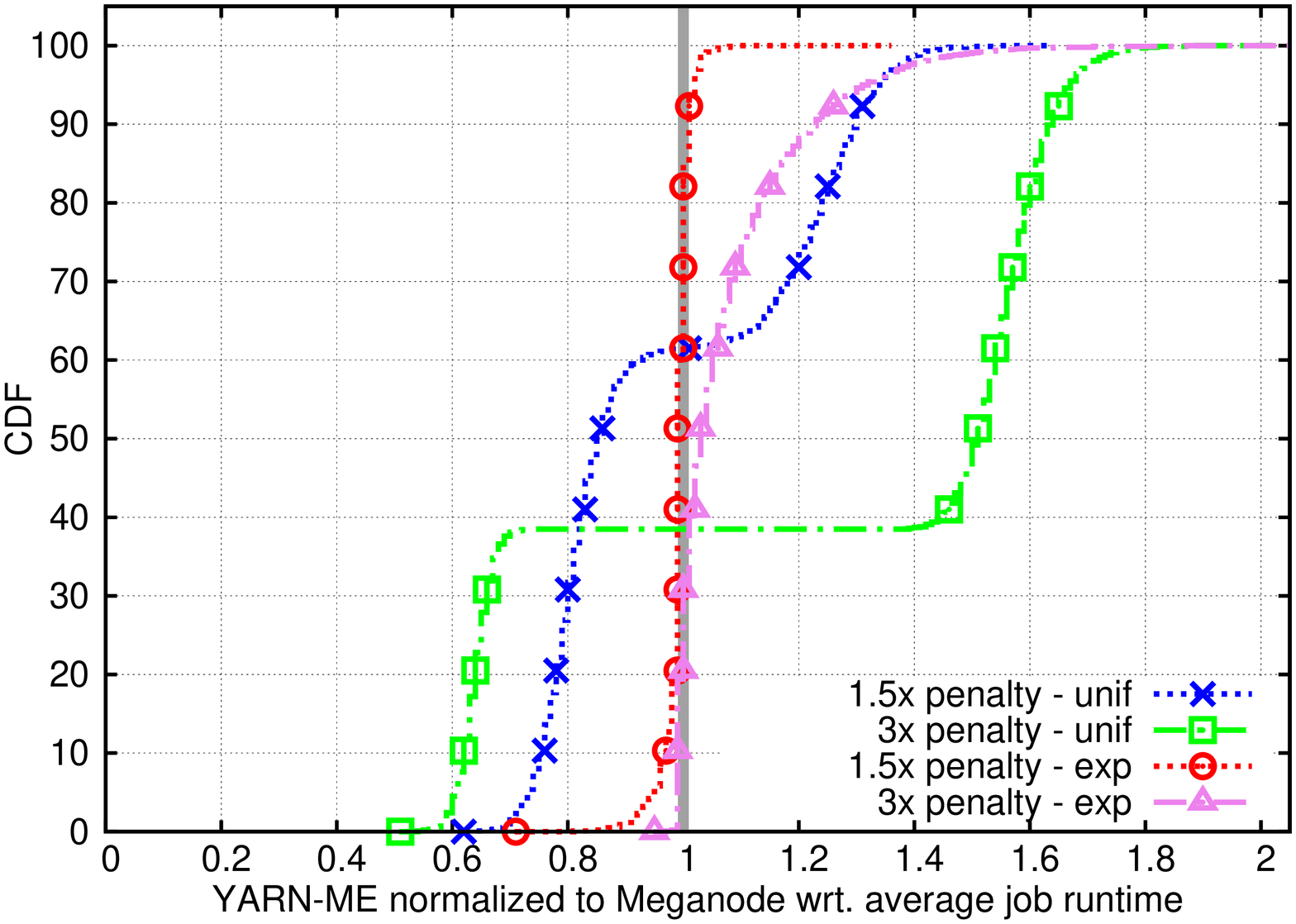}}
	\caption{CDFs of average job runtime in YARN-MEs vs YARN (a) and vs the idealized Meganode (c). YARN's benefits hold for traces of varying size (b).}
	\vspace{-0.2in}
\end{figure*}

\subsection{Simulation Methodology}

{\bf Simulator\hspace{0.2in}} 
We built DSS (Discrete Scheduler Simulator) a discrete-time simulator for YARN
and we made the code publicly
available~\footnote{https://github.com/epfl-labos/DSS}. In DSS, simulated tasks
do not perform computation or I/O. The tasks are simulated using task start and
task finish events. We simulate a cluster with 16 cores and 10GB of RAM per
node. Memory is assigned to tasks with a granularity of 100MB.  Jobs are
ordered according to YARN's FairScheduling~\cite{yarn-fairscheduler} policy.
Each task uses 1 core. The minimum amount of memory allocatable to a task is
set to 10\% of its ideal requirement. We use a 100-node cluster to perform a
parameter sweep but also show results for up to 3000 nodes.

\noindent{\bf Simulation traces\hspace{0.2in}}
A trace contains for each job: the job submission time, the number of tasks,
the ideal amount of memory for a task and task duration at the ideal amount of
memory. Each job has one parallel phase. Job arrivals are uniformly random
between 0 and 1000s. The other parameters are varied according to either a
uniform or an exponential random distribution. We use 100-job traces but also
show results for up to 3000 jobs.

\noindent{\bf Modeling elasticity\hspace{0.2in}}
Since the simulated jobs have only one single phase we only use the reducer
penalty model from \S\ref{sec:study}. We show results for 1.5x and
3x penalties, to cover the range of penalties measured in \S\ref{sec:study}.

\noindent{\bf Metrics\hspace{0.2in}} We use average job runtime to compare the
different approaches.

\subsection{Simulation experiments}

\noindent{\bf YARN-ME vs YARN\hspace{0.2in}} We perform a parameter sweep on 3
trace parameters: memory per task, tasks per job and task duration. The table
below shows the different parameter ranges. We keep the min constant and vary
the max within an interval to perform the sweep. This gives us a range for each
parameter, which is then varied independently of the others. We draw the values
from a uniform or exponential random distribution. We perform 100 runs for each
combination of 3 ranges (one for each parameter) and show the median and the
maximum (worst-case) results for normalizing YARN-ME to YARN in
\cref{fig:ELASTIC_vs_YARN_jrt}. We use 100-job traces on 100 nodes.

\vspace{-0.05in}
\begin{center}
\scalebox{0.88}{
\begin{tabular}{ |c|c|c|c|c|c|c| }
\hline
\multirow{3}{*}{dist} & \multicolumn{2}{c|}{tasks / job} & \multicolumn{2}{c|}{mem / task} & \multicolumn{2}{c|}{task duration} \\
\cline{2-3}
& \multirow{2}{*}{min} & \multirow{2}{*}{max} & \multicolumn{2}{c|}{(GB)} & \multicolumn{2}{c|}{(s)} \\
\cline{4-7}
& & & min & max & min & max \\
\hline
unif & 1 & $[200,400]$ & 1 & $[2,10]$ & 1 & $[200,500]$ \\
\hline 
exp & 1 & $[20,220$] & 1 & $[2,10]$ & 50 & $[100,500]$ \\
\hline
\end{tabular}}
\end{center}

The uniform distribution yields bigger benefits because it leads to more memory
fragmentation in YARN.  As expected, YARN-ME's improvements are larger if
penalties are lower. The case in which YARN-ME does not improve on YARN are
either cases where the cluster utilization is very low or when most tasks have
very small memory requirements. In such cases, memory elasticity is less
beneficial.  Nevertheless, for 3x penalty and uniform distribution, 40\% of the
configurations have a ratio of YARN-ME to YARN of at most 0.7. 

\cref{fig:ELASTIC_vs_YARN_largetrace} shows the behavior of one uniform
trace in a weak scaling experiment. We scale the trace and cluster size
simultaneously from 100 to 3000. The benefits of YARN-ME hold despite the
scaling.

\noindent{\bf The need for elasticity (YARN-ME vs Meganode)\hspace{0.2in}} We
next show that YARN-ME yields improvements beyond the reach of current
elasticity-agnostic schedulers. We compare against an idealized scenario
(called Meganode) which serves as an upper-bound for elasticity-agnostic
solutions that improve average job runtime. The Meganode pools all cluster
resources into one large node with a memory and core capacity equal to the
aggregate cluster-wide core and memory capacity available for YARN-ME.  Thus,
the Meganode does away with machine-level memory fragmentation. Meganode uses a
shortest remaining job first (SRJF) scheduling policy because that is known to
improve average job runtime. However, in using SRJF, the Meganode loses all
fairness properties whereas YARN-ME obeys the existing fairness policy.

\cref{fig:ELASTIC_vs_meganode_jrt} compares the averge job runtime for
Meganode and YARN-ME on 20.000, 100-job traces on 100 nodes. While it is
expected that Meganode wins in many cases, YARN-ME beats Meganode for 40\%-60\%
of the cases for the uniform trace and for 20\% of the exponential trace for
1.5x penalty. YARN-ME gains because it turns even small amounts of memory
fragmentation into an opportunity by scheduling elastic tasks.

\begin{figure}[t]
    \includegraphics[width=2.2in,angle=270]{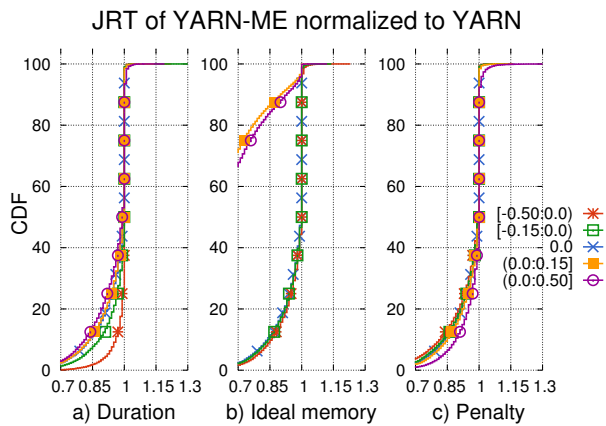}
    \caption{Sensitivity to mis-estimations. 3x penalty.}
    \label{fig:misest}
    \vspace{0.1in}
\end{figure}

\noindent{\bf Sensitivity to mis-estimations\hspace{0.2in}}
Further, we show that YARN-ME is robust to mis-estimations. We generate 20,000
traces with each of the trace parameters (memory per task, tasks per job, and
task duration) following an exponential random distribution, within bounds of
$[0.1, 10]$ GBs, $[1, 100]$ tasks, and $[50, 500]$ seconds. We simulate
mis-estimations by altering the duration, ideal memory, and performance penalty
of tasks for both regular and elastic allocations. This forces the scheduler to
make decisions based on imperfect information. We change each parameter
fractionally by a uniformly random factor in the intervals of $(0, 0.15]$, and
$(0, 0.5]$ (0.15 represents a 15\% higher value). The former interval
represents the worst-case deviation of our model in
\cref{fig:hadoop-reducer-model}, while the latter is an extreme example chosen
to stress YARN-ME. We present both positive and negative mis-estimations.
\cref{fig:misest} presents the ratio between average job completion time with
YARN-ME and YARN, for an elasticity penalty of 3x -- one of the highest
penalties measured in \S\ref{sec:study}.

\noindent{\bf Sensitivity to task duration mis-estimation\hspace{0.2in}}
YARN-ME is robust to task duration mis-estimation, which can occur due to system
induced stragglers or data locality. The timeline generator of the simulator
bases its information on task durations from the trace. We alter each actual
task runtime by a different factor.

For $[-0.15, 0.5]$, YARN-ME achieves gains similar to the scenario without
mis-estimations on all traces. Even for the very large $[-0.5, 0)$
mis-estimations, the gains are still comparable, with only $\sim$35\% of the
traces reporting at most 10\% lower gains. This is due to tasks being shorter
than the timeline generator expects, resulting in a small number of elastic
tasks exceeding the estimated job completion time.

\noindent{\bf Sensitivity to model mis-estimations\hspace{0.2in}}
Finally, YARN-ME is also robust to model mis-estimations, which may occur during
profiling. We change task memory (\cref{fig:misest}b) and penalty
(\cref{fig:misest}c) with a different value for each job.

YARN-ME improves by up to 45\% in the case of positive mis-estimation of ideal
memory (\cref{fig:misest}b). In this case, all tasks (in both YARN and YARN-ME)
spill data to disk and become penalized tasks. However, penalties in YARN-ME are
lower because YARN-ME can choose the under-sized allocation that minimizes
penalty while YARN lacks this capability. Negative mis-estimation of ideal
memory has negligible impact.

In the case of penalty mis-estimation (\cref{fig:misest}c), only the $(0, 0.5]$
runs exhibit gains reduced by at most 4\%. This is due to the scheduler being
more conservative since it perceives elastic tasks as taking longer to execute.

\section{Related Work}
\label{sec:relatedwork}

Current schedulers do not leverage memory elasticity. Next, we review the
mechanisms from current schedulers that are most related in spirit to memory
elasticity.

Tetris~\cite{tetris} improves resource utilization (including memory) by better
packing tasks on nodes. It adapts heuristics for the multi-dimensional bin
packing problem to the context of cluster scheduling. However it estimates a
task's peak memory requirements and only schedules the task on a node that has
enough memory available to cover the peak.

Heracles~\cite{heracles} aggressively but safely collocates best-effort tasks
alongside a latency critical service. It does this by dynamically managing
multiple hardware and software mechanisms including memory. However, Heracles
only considers RAM bandwidth and not capacity.

Apollo~\cite{apollo} is a distributed scheduler that provides an
opportunistic scheduling mode in which low priority tasks can be scheduled
using left-over memory unused by normal priority tasks. Normal priority tasks
are scheduled only if their resource demands are strictly met.  Apollo has no
principled way of reasoning about the performance implications of opportunistic
allocations nor does it provide a decision mechanism about when such
allocations are useful. Borg~\cite{borg} provides similar capabilities with a
centralized design.

Quasar~\cite{quasar} leverages machine-learning classification techniques to
reason about application performance with respect to scale-up allocations.  A
greedy algorithm places tasks starting with nodes that give the best
performance satisfying application SLOs and improving resource utilization.
Quasar does not identify or discuss memory elasticity.

ITask~\cite{itask-sosp} is a new type of data-parallel task that can be
interrupted upon memory pressure and have its memory reclaimed. The task can
then be resumed when the pressure goes away. ITask is a system-level mechanism
that uses preemption to mitigate unmanageable memory pressure before it can hurt
system performance. Memory elasticity can work in tandem with ITask, since
elastic tasks will need less time to spill, and thus can be preempted and
resumed faster than regular tasks.

\section{Conclusion}
\label{sec:conclusion}

The main contribution of this paper is identifying, quantifying and
demonstrating memory elasticity, an intrinsic property of data-parallel
workloads. Memory elasticity allows tasks to run with significantly less memory
than ideal while incurring only a moderate performance penalty.  We show that
memory elasticity is prevalent in the Hadoop, Spark, Tez and Flink frameworks.
We also show its predictable nature by building simple models for Hadoop and
extending them to Tez and Spark. Applied to cluster scheduling, memory
elasticity helps reduce task completion time by decreasing task waiting time
for memory. We show that this can be transformed into improvements in job
completion time and cluster-wide memory utilization. We integrated memory
elasticity into Apache YARN and showed up to 60\% improvement in average job
completion time on a 50-node cluster running Hadoop workloads.

\bibliography{ms}
\bibliographystyle{abbrv}

\end{document}